\documentclass[11pt]{article}
\usepackage{mathrsfs}
\usepackage{amsmath}
\usepackage{amssymb}
\usepackage{amsfonts}
\usepackage{CJK}
\usepackage{subfigure}
\usepackage{graphicx}
\usepackage{dsfont}
\usepackage{amscd}
\usepackage{pgfplots}
\usepackage[body={16.5cm,24cm}, top=1cm]{geometry}
\newtheorem{thm}{Theorem}

 \newtheorem{lem}{Lemma}
 \newtheorem{prop}{Proposition}

\begin{document}

\title{Modulational instability and homoclinic orbit solutions in vector nonlinear Schr\"odinger equation}
\author{Liming Ling$^1$, Li-Chen Zhao$^{2,3}$ \\
\\$^1$School of Mathematics, South China University of Technology, Guangzhou 510640, China\\
$^2$ School of Physics, Northwest University,  Xi'an  710069, China\\
$^3$ Shaanxi Key Laboratory for Theoretical Physics Frontiers, Xi'an, 710069, China}

\maketitle
\begin{abstract}
Modulational instability has been used to explain the formation of breather and rogue waves qualitatively. In this paper, we show modulational instability can be used to explain the structure of them in a quantitative way. We develop a method to derive general forms for Akhmediev breather and rogue wave solutions in a $N$-component nonlinear Schr\"odinger equations. The existence condition for each pattern is clarified clearly. Moreover, the general multi-high-order rogue wave solutions and multi-Akhmediev breather solutions for $N$-component nonlinear Schr\"odinger equations are constructed. The results further deepen our understanding on the quantitative relations between modulational instability and homoclinic orbits solutions.

\

\textbf{Key words:} Akhmediev breathers, general multi-high-order rogue wave, modulational instability analysis, vector nonlinear Schr\"odinger equation.

MSC2010: {37K10,35Q55,35C08}
\
\end{abstract}

\maketitle

\section{Introduction}
Akhmediev breathers (ABs) and Rogue waves (RWs) including scalar ones and vector ones, have been observed in many
different physical systems \cite{RWRep,Kibler,Chabchoub,Bailung,Frisquet2}. RWs are found to have many different fundamental patterns, such as eye-shaped
one, anti-eye-shaped one, and four-petaled one, etc. Scalar ones usually admit eye-shaped
one \cite{Kharif,Pelinovsky}, while vector ones admit
anti-eye-shaped one and four-petaled one  \cite{Guo,Bludov,zhao2,Chen,Zhao3,Degasperis,Chen3}. ABs with infinite period tend to be RW, but ABs with finite period admit different pattern with RW (the hump amplification rate is different and the valley values are also distinctive). The structure of AB can be characterized by its period and the localized wave structure in one period (be seen as a unit of AB). Since the dynamics of ABs is approaching to plane wave background when $t$ tends to $\pm\infty$ and the dynamics of RWs is approaching to plane wave background when $x,t$ tend to $\pm\infty$, we call them as the homoclinic orbit solutions\cite{MacL}. On the other hand, since ABs are localized in time and periodic in space and RWs are localized both in time and space, then they also belong to the localized wave solutions. The nonlinear superpositions of these fundamental localized waves correspond to high-order ones or multiple ones, which could admit much more complex structures. Therefore, the understanding of fundamental ones is the first step to explain complex dynamics of AB or RW related processes. The well-known ABs and RWs solutions are related with the MI analysis. The relation between ABs and MI in the coupled NLSE was uncovered in ten years ago.
Recently, there are some studies to consider the relations between RWs and MI in the nonlinear Schr\"odinger type equations
\cite{Baronio1,zhaoling,NMI}.

To the best of our knowledge, there still exist some unclear problems in the VNLSE model.
Firstly,
the AB solutions for the vector NLSE (VNLSE) are constructed by the Darboux transformation. The single AB solutions for the VNLSE were well known for us more than a decade years ago. With the aid of the Darboux transformation, the multi-AB solution also can be constructed. But if we do not find the proper parameter, the distinct AB possesses different period and the multi-AB can not be periodical solution but the quasi-periodical solution which can not be explained by MI. A way to derive the multi-AB with the same period had not been involved in the previous research.
Secondly,
the parameter $\lambda$ in the Lax pair is the spectral parameter which is related with the inverse scattering method. For the classical NLSE, the spectral points which correspond to the AB solutions can be solved exactly by the twice algebraic equation. For the multi-component system which correspond the high order system, the root of the high-order algebraic equation can not be solved with a compact form which hinders us to analyze the spectral points where the ABs can be obtained. Thirdly,
the MI analysis with the finite period $T$ can be used to explain the amplifying process of ABs. While the periodical parameter $T$ approaches to the infinite, MI could be used to illustrate the amplifying process of RWs. However, the qualitative relations between MI and homoclinic orbits solutions have not been exposed. Meanwhile, besides the fundamental ones, there still exist another homoclinic orbits solutions that could be accounted for the MI.

In this paper, we focus on the aforementioned problems. We arrange the paper as follows. In section \ref{sec2}, we present a method to derive general forms of localized wave on plane wave background through the Darboux dressing method.  In section \ref{sec3}, we present fundamental ABs solutions and multi-AB solutions with the same period. Moreover, we discuss the relation between the characters of spectral parameter and AB structure. The two-component model is analyzed in detail. In section \ref{sec4}, we derive general RW solution, multi-RW, and even high-order multi-RW solutions, and characterize the structure of fundamental RW. Here we should point out that the construction for multi-high-order RWs solutions is firstly obtained in this work. In section \ref{sec5}, we explain the structure of fundamental ones based on linear stability analysis. We uncover that the patterns are determined by the dispersion form of weak perturbations on plane wave backgrounds. In section \ref{sec6},  we test the prediction of AB or RW pattern based on MI analysis numerically. Finally, we summarize the results and present some discussions on future works in section \ref{sec7}.

\section{A method to derive localized wave solutions on plane wave background of VNLSE}\label{sec2}
Until now, most of previous studies on RWs and ABs were focused on low-component systems such as one-, two- or three-component systems \cite{Bludov,zhao2,Chen,Zhao3,Degasperis}. However, in real physical systems, there are many coupled systems which have much more components, such as  five- (spin-2 spinor Bose-Einstein condensate \cite{Spin2}) or seven-component ones (spin-3 spinor Bose-Einstein condensate \cite{Spin3}), and multi-mode fiber \cite{multi-mode}. In comparison with low-component systems,  the dynamical behaviors
and relevant patterns of RWs and ABs of high-component systems are less studied, partly because of the difficulties in solving $N$-component coupled nonlinear equations on nonzero backgrounds. We would like to present general solution forms for fundamental RWs (ABs) and nonlinear superpositions of them, which can be used to investigate their dynamics analytically and exactly.

We consider the following focusing vector nonlinear
Schr\"odinger equation (VNLSE)
\begin{equation}\label{CNLSE}
    {\rm i}\mathbf{q}_t+\frac{1}{2}\mathbf{q}_{xx}+\mathbf{q}\mathbf{q}^{\dag}\mathbf{q}=0,
\end{equation}
where $\mathbf{q}=\left(q_1,q_2,\cdots,q_N\right)^T$, ``$\dag$" represents the Hermite conjugation. The VNLSE can be used to describe evolution of localized waves in a  nonlinear fiber with two modes \cite{Agwal}, two-component Bose-Einstein condensate \cite{Spinor}, and other coupled nonlinear systems \cite{Guo,Baronio}. It admits the following Lax pair:
\begin{equation}\label{Lax}
    \begin{split}
       \Phi_x &=U(\lambda;Q)\Phi, \\
       \Phi_t &=V(\lambda;Q)\Phi,\\
    \end{split}
\end{equation}
where
\begin{equation*}
\begin{split}
    U(\lambda;Q)&={\rm i}\left[\frac{\lambda}{2} (\sigma_3+I_{N+1})+Q\right],  \\
    V(\lambda;Q)&={\rm i}\left[\frac{\lambda^2}{4}(\sigma_3+I_{N+1})+\frac{\lambda}{2} Q-\frac{1}{2}\sigma_3(Q^2+{\rm i}Q_x)+|a|^2 I_{N+1}\right] \\
   Q &=\begin{pmatrix}
        0 & \mathbf{q}^{\dag} \\
        \mathbf{q} & 0  \\
      \end{pmatrix}
    ,\,\left.
         \begin{array}{ll}
           \sigma_3&=\mathrm{diag}(1,-I_N),
         \end{array}
       \right.
\end{split}
\end{equation*}
$I_k$ is a $k\times k$ identity matrix, the parameter $|a|^2=\sum_{i=1}^{N}|a_i|^2$ is a real constant, $\lambda\in \mathbf{C}\cup {\infty}$ is a complex parameter. The compatibility condition $\Phi_{xt}=\Phi_{tx}$ gives the vector NLSE \eqref{CNLSE}.

We can convert the system \eqref{Lax} into a new system
\begin{equation}\label{CNLSE1}
    \begin{split}
   \Phi[1]_x &=U(\lambda;Q[1])\Phi[1],\\
       \Phi[1]_t &=V(\lambda;Q[1])\Phi[1],
        \end{split}
\end{equation}
 by the following elementary
Darboux transformation \cite{Matveev,Gu}
\begin{equation}\label{DT}\begin{split}
                             \Phi[1]&=T\Phi,\,T=I+\frac{\lambda_1^*-\lambda_1}{\lambda-\lambda_1^*}P_1,\,\, P_1=\frac{\Phi_1\Phi_1^{\dag}}{\Phi_1^{\dag}\Phi_1}, \\
                              \mathbf{q}[1]&=\mathbf{q}+\frac{(\lambda_1^*-\lambda_1)\phi_1^*}{|\phi_1|^2+\sum_{i=1}
                              ^{N}|\psi_1^{[i]}|^2}\mathbf{\psi}_1,
                          \end{split}
\end{equation}
where $\Phi_1=(\phi_1,\mathbf{\psi}_1^T)^T,$ is a special solution for system \eqref{CNLSE} at $\lambda=\lambda_1$, and $\mathbf{\psi}_1=D_N(\psi_1^{[1]},\psi_1^{[2]},\cdots,\psi_1^{[N]})^T$, $D_N=\left(a_1{\rm e}^{\theta_1},\,a_2{\rm e}^{\theta_2},\cdots,\, a_N{\rm e}^{\theta_N}\right)$.

\begin{thm}
The $n$-fold Darboux transformation can be represented as
\begin{equation}\label{nDT}
  T_n=I_n+YM^{-1}(\lambda I_n-D)^{-1}Y^{\dag},
\end{equation}
where
\begin{equation*}
  \begin{split}
      Y=&\left[|y_1\rangle,|y_2\rangle,\cdots, |y_n\rangle\right]=\begin{bmatrix}
                                                                    Y_1 \\
                                                                    Y_2 \\
                                                                  \end{bmatrix}
      ,  \\
      D=&\mathrm{diag}\left(\lambda_1,\lambda_2,\cdots,\lambda_n\right),\\
      M=&\left(\frac{\langle y_i|y_j\rangle}{\lambda_i^*-\lambda_j}\right)_{1\leq i,j\leq n},
  \end{split}
\end{equation*}
$|y_i\rangle$ is a special solution for Lax pair equation \eqref{Lax} at $\lambda=\lambda_i$, $Y_1$ is a $1\times n$ matrix, $Y_2$ is a $N\times n$ matrix.
And the B\"acklund transformations between old potential functions and new ones are
\begin{equation}\label{back-n}
  \mathbf{q}[n]=\mathbf{q}+Y_2M^{-1}Y_1^{\dag}.
\end{equation}

\end{thm}
It should be pointed out that bright soliton and high-order soliton or multi-soliton have been derived systemically \cite{vs,vs2,vs3,feng}. However, the localized waves on plane wave backgrounds have not been obtained for VNLSE with arbitrary component number, since it is hard to solve related high-order algbraic equation.
To obtain the general localized wave solutions for VNLSE, we choose a general plane wave solution
\begin{equation}\label{seed}
    q_i[0]=a_i \exp{\theta_i},
\end{equation}
where $\theta_i={\rm i} \left [b_i x + (|a|^2-\frac{1}{2}b_i^2) t \right]$, $i=1,2,\cdots,N$.
We firstly investigate fundamental solution of Lax-pair with the plane wave solution to develop a new method for deriving nonlinear wave solutions for $N$ components case. Substituting seed solution \eqref{seed} into equation
\eqref{Lax}, we can obtain a vector solution of the Lax pair
\begin{equation*}
    \Phi_{j,s}=D_{N+1}\begin{bmatrix}
             \exp{\omega_{j,s}}
             \\[4pt]
             \displaystyle{\frac{\exp{\omega_{j,s}}}{\chi_{j,s}+b_1}} \\[8pt]
             \vdots \\
             \displaystyle{\frac{\exp{\omega_{j,s}}}{\chi_{j,s}+b_N}}
           \end{bmatrix}
\end{equation*}
where
\begin{equation*}
    \begin{split}
      D_{N+1}&=\mathrm{diag}\left(1,D_N\right), \\
      \omega_{j,s}&={\rm i}\chi_{j,s}\left( x+\frac{1}{2}\chi_{j,s}t+\vartheta_{j,s}\right)
    \end{split}
\end{equation*}
$\vartheta_{j,s}$ is a complex parameter and $\chi_{j,l}$ is a root for the following $N+1-$th order algbraic equation
\begin{equation}\label{cubic}
   \prod_{i=1}^{N}(\chi+b_i)\left[\chi-\lambda_j-\sum_{i=1}^{N}\frac{a_i^2}{\chi+b_i}\right]=0.
\end{equation}

\begin{prop}\label{prop1}
If the following identities
\begin{equation*}
\begin{split}
    & \chi_{j,s}-\lambda_j-\sum_{i=1}^{N}\frac{a_i^2}{\chi_{j,s}+b_i}=0, \\
    & \chi_{k,l}-\lambda_k-\sum_{i=1}^{N}\frac{a_i^2}{\chi_{k,l}+b_i}=0,
\end{split}
\end{equation*}
are valid,
then we can obtain that
\begin{equation*}
  \frac{\lambda_k^*-\lambda_j}{\chi_{k,l}^*-\chi_{j,s}}=
1+\sum_{i=1}^{N}\frac{a_i^2}{(\chi_{j,s}+b_i)(\chi_{k,l}^*+b_i)}.
\end{equation*}
\end{prop}

Based on above proposition and linear algebra, we obtain the following nonlinear wave solution
\begin{equation}\label{gene-soliton}
  q_i[K]=a_i\left[\frac{\det(M_i)}{\det(M)}\right]\exp(\theta_i),
\end{equation}
where
\begin{equation*}
  \begin{split}
      M=&(m_{k,j})_{1\leq k,j\leq K}  \\
      M_i=&(m_{k,j}^{i})_{1\leq k,j\leq K}
  \end{split}
\end{equation*}
and
\begin{equation*}
  \begin{split}
     m_{k,j}=&\sum_{l=1,s=1}^{N,N}\frac{\delta_{k,l}\delta_{j,s}\exp(\omega_{k,l}^*+\omega_{j,s})}{\chi_{j,s}-\chi_{k,l}^*},  \\
     m_{k,j}^{i}=&\sum_{l=1,s=1}^{N,N}\frac{\chi_{j,s}^*+b_i}{\chi_{k,l}+b_i}\frac{\delta_{k,l}\delta_{j,s}\exp(\omega_{k,l}^*+\omega_{j,s})}{\chi_{j,s}^*-\chi_{k,l}},
  \end{split}
\end{equation*}
$\delta_{k,l}=0,1.$
This is a general form which can be used to construct many different types of nonlinear localized waves on plane wave backgrounds. In this paper, we just apply them to construct AB and RW solution explicitly, and try to explain the mechanism of their spatial-temporal structures.

\section{The general Akhmediev breather solutions of VNLSE}\label{sec3}
If two roots possess an identical imaginary part, then AB solution can be
constructed. Aussume a root $\chi$ satisfies
\begin{equation}\label{charaN1}
  \chi-\lambda-\sum_{i=1}^{N}\frac{a_i^2}{\chi+b_i}=0.
\end{equation}
Another root is denoted as $\chi+\alpha$, which satisfies
\begin{equation}\label{charaN2}
  \chi+\alpha-\lambda-\sum_{i=1}^{N}\frac{a_i^2}{\chi+\alpha+b_i}=0.
\end{equation}
Equation \eqref{charaN1} subtracts \eqref{charaN2}, it follows that
\begin{equation}\label{charaN3}
  1+\sum_{i=1}^{N}\frac{a_i^2}{(\chi+b_i)(\chi+\alpha+b_i)}=0,
\end{equation}
which is nothing but the determined equation for AB solution.

Then, the fundamental AB solutions with the period $2\pi/\alpha$ in $x$-direction can be represented as
\begin{equation}\label{ele}
  q_i=a_i\left(\frac{{\displaystyle \frac{\chi_1^*+b_i}{\chi_1+b_i}{\rm e}^{\omega_R}+
  \frac{\chi_1^*+b_i+\alpha}{\chi_1+b_i+\alpha}{\rm e}^{-\omega_R}+\frac{\chi_1^*+b_i}{\chi_1+b_i+\alpha}\frac{2\chi_{1{\rm i}}{\rm e}^{\omega_I}}{2\chi_{1{\rm i}}-{\rm i}\alpha}+\frac{\chi_1^*+b_i+\alpha}{\chi_1+b_i}\frac{2\chi_{1{\rm i}}{\rm e}^{-\omega_I}}{2\chi_{1{\rm i}}+{\rm i}\alpha}}}{{\displaystyle 2\cosh(\omega_R)+\frac{2\chi_{1{\rm i}}{\rm e}^{\omega_I}}{2\chi_{1{\rm i}}-{\rm i}\alpha}+\frac{2\chi_{1{\rm i}}{\rm e}^{-\omega_I}}{2\chi_{1{\rm i}}+{\rm i}\alpha}}}\right){\rm e}^{\theta_i}
\end{equation}
through the formula \eqref{gene-soliton} by choosing proper parameters, where $\omega_I={\rm i}\alpha\left[x+\left(\chi_{1{\rm r}}+\frac{\alpha}{2}\right)t\right]$, $\omega_R=\alpha\chi_{1{\rm i}}t$, $\chi_1=\chi_{1{\rm r}}+\chi_{1{\rm i}}{\rm i}$. Without loss of generality, we assume that $\alpha,\mathrm{Im}(\chi_{1})>0$. Then, the equation \eqref{charaN3} determines the structure of fundamental AB unit and period uniquely. We choose a two-component case to discuss the dynamics of AB explicitly, since the solution can be written in a more precise and simpler form in this case.

The shape of AB can be described by the value of $|q_i|$ located on the center $(x,t)=(x_0,t_0)$,$x_0=\frac{(2k+1)\pi}{\alpha}-\arccos\frac{2\chi_{1{\rm i}}}{\sqrt{\alpha^2+4
\chi_{1{\rm i}}^2}}$, $t_0=0$, $k\in \mathbb{Z}$,
\begin{equation*}
  |q_i(x_0,t_0)|=a_i^2\left|\frac{\frac{\alpha^{2}\delta}{2(1-\delta)}-\left[\left(\chi_{1{\rm r}}+b_{i} \right) \alpha+(\chi_{1{\rm r}}+b_{i})^2+\chi_{1{\rm i}}^{2} \right]}{|(\chi_1+b_i)(\chi_1+\alpha+b_i)|}\right|,
\end{equation*}
where $\delta=2\chi_{1{\rm i}}/\sqrt{\alpha^2+4\chi_{1{\rm i}}^2}$. We can classify the dynamics through the above value of $|q_i(x_0,t_0)|$ which depends on the variables $\chi_1$, $\alpha$ and $b_i$. Since it is rather complicated, here we omit to analyze the explicit dynamics through this value in detail.

\subsection{Akhmediev breather classification in a two-component case}

In this paragraph, we consider AB solutions in a two-component case, which is temporal periodic and spatial localized in the whole $(x,t)$ plane. Through the above general theory, we obtain the determined conditions for AB solution:
\begin{equation}\label{chara3}
  1+\sum_{i=1}^{2}\frac{a_i^2}{(\chi+b_i)(\chi+\alpha+b_i)}=0.
\end{equation}
For fixed parameters $a_i,b_i$ ($i=1,2$),  one can determine the spectral point $\lambda$
for which there are two roots with the same image part of \eqref{chara3}. To achieve this aim, we firstly
solve the equation \eqref{chara3}. In general, the quantic equation can be solved with formula \eqref{chara3}, but it is rather so complex that we can not analyze it conveniently. Since the equation possesses the Galieo transformation, then we can suppose $b_1=-b_2=\beta>0$.
From the above complex roots classification condition with $\beta$, we further find that there \textbf{are mainly three cases for the numbers of complex root}. If $\alpha=0$, there are two pairs of conjugated complex roots. If $|\alpha|>2|\beta|$, we deduce that there are at least two real roots for equation \eqref{chara3}, which locate on the interval $(-\beta,\beta)$ and $(-(\beta+\alpha), \beta-\alpha)$ respectively. Therefore, we infer that the two pairs of conjugated complex roots are merely possible to exist in the interval $|\alpha|\in(0,2|\beta|)$.  For different complex root numbers, AB dynamics in the two components demonstrate distinctive behaviors.

For a particular case $a_1=a_2=1$, we can obtain a simple formula for the equation \eqref{chara3}:
\begin{equation}\label{equation1}
  \begin{split}
     \chi_{1,\pm}=&-\frac{\alpha}{2}\pm\frac{1}{2}\sqrt {{\alpha}^{2}+4{\beta}^{2}-4+4\sqrt {({\alpha}^{2}-4){
\beta}^{2}+1}},
  \\
     \chi_{2,\pm}=&-\frac{\alpha}{2}\pm\frac{1}{2}\sqrt {{\alpha}^{2}+4{\beta}^{2}-4-4\sqrt {({\alpha}^{2}-4){
\beta}^{2}+1}}.
  \end{split}
\end{equation}
It is readily to see that the conditions $({\alpha}^{2}-4){
\beta}^{2}+1<0$ or $E_{\pm}\equiv{\alpha}^{2}+4{\beta}^{2}-4\pm4\sqrt {({\alpha}^{2}-4){
\beta}^{2}+1}<0$ can ensure that $\chi_{1,\pm}$ and $\chi_{2,\pm}$ are complex numbers.
Then we conclude that
\begin{itemize}
  \item  $0<\beta<1/2$: When $-2\beta\leq \alpha\leq 2\beta$, then $\chi_{1,\pm}$ are complex roots; while
$-2\sqrt{2+\beta^2}\leq \alpha\leq 2\sqrt{2+\beta^2}$, $\chi_{2,\pm}$ are complex roots.
  \item $1/2\leq \beta \leq \sqrt{2}/2$: While $-\sqrt{4-\beta^{-2}}\leq \alpha\leq \sqrt{4-\beta^{-2}}$ or $\sqrt{4-\beta^{-2}}\leq |\alpha|\leq 2\beta$, then $\chi_{1,\pm}$ are complex roots; when
$-\sqrt{4-\beta^{-2}}\leq \alpha\leq \sqrt{4-\beta^{-2}}$ or $\sqrt{4-\beta^{-2}}\leq |\alpha|\leq 2\sqrt{2+\beta^2}$, $\chi_{2,\pm}$ are complex roots.
  \item $\beta>\sqrt{2}/2$: When $-\sqrt{4-\beta^{-2}}\leq \alpha\leq \sqrt{4-\beta^{-2}}$, then $\chi_{1,\pm}$ are complex roots; while
$-\sqrt{4-\beta^{-2}}\leq \alpha\leq \sqrt{4-\beta^{-2}}$ or $2\beta\leq |\alpha|\leq 2\sqrt{2+\beta^2}$, $\chi_{2,\pm}$ are complex roots.
\end{itemize}

\begin{figure}[tbh]
\centering
\subfigure[$\beta=\frac{1}{4}$]{%
\includegraphics[height=50mm,width=65mm]{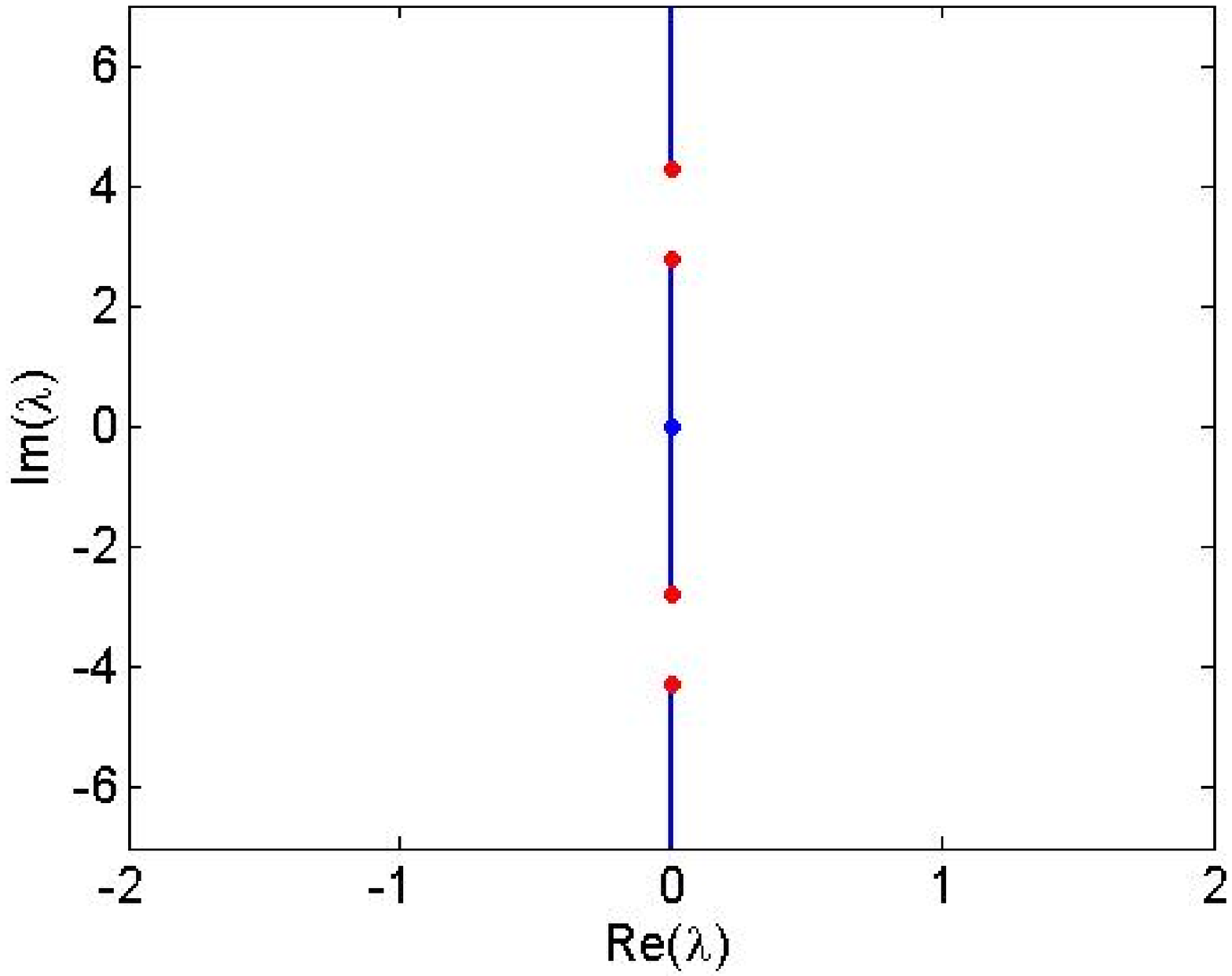}} \hfil
\subfigure[$\beta=\frac{1}{2}$]{%
\includegraphics[height=50mm,width=65mm]{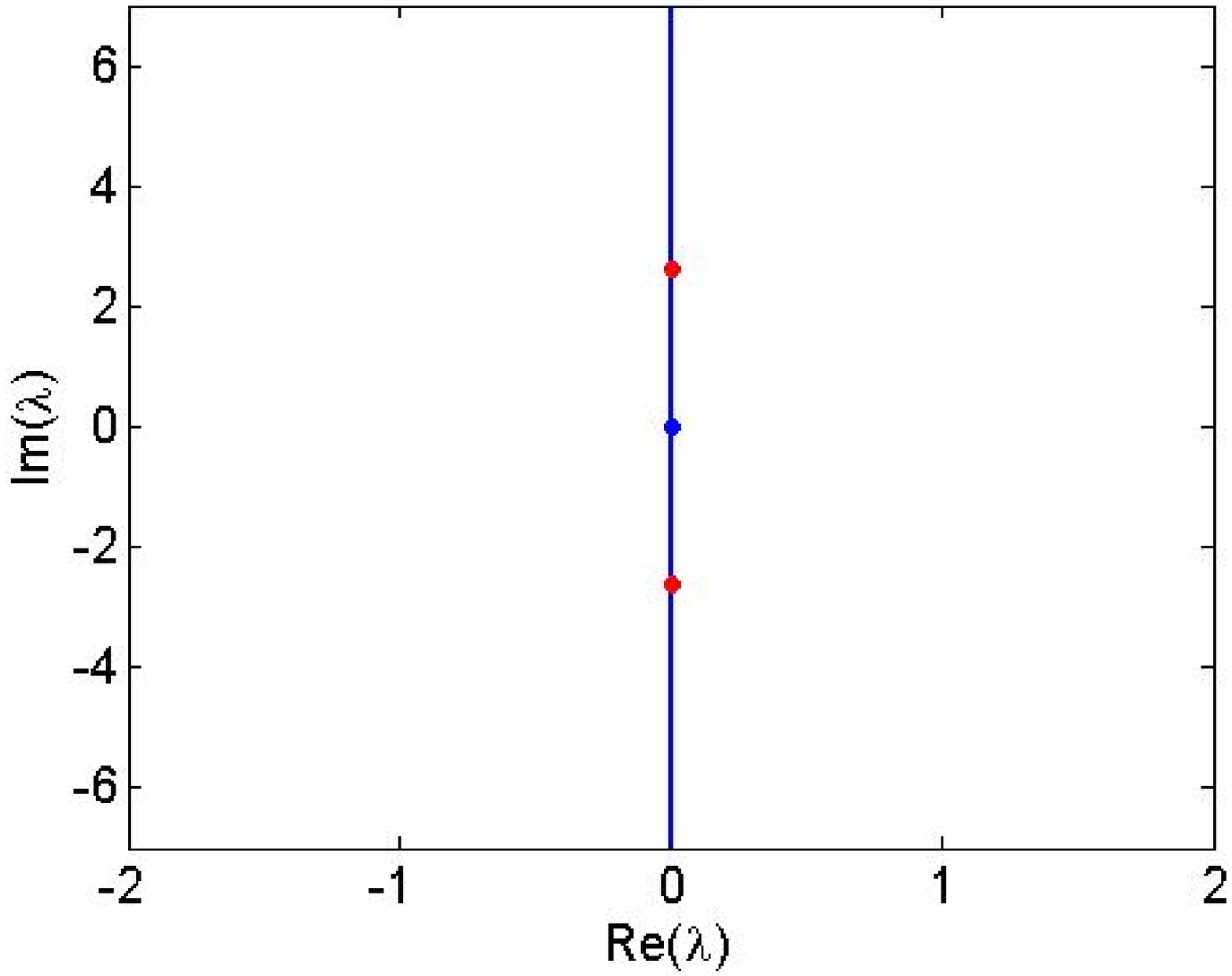}}
\caption{(color online): The red points represent the branch points for spectral curve $\lambda$.}
\label{con1}
\end{figure}

\begin{figure}[tbh]
\centering
\subfigure[$\beta=\frac{7}{10}$]{%
\includegraphics[height=50mm,width=65mm]{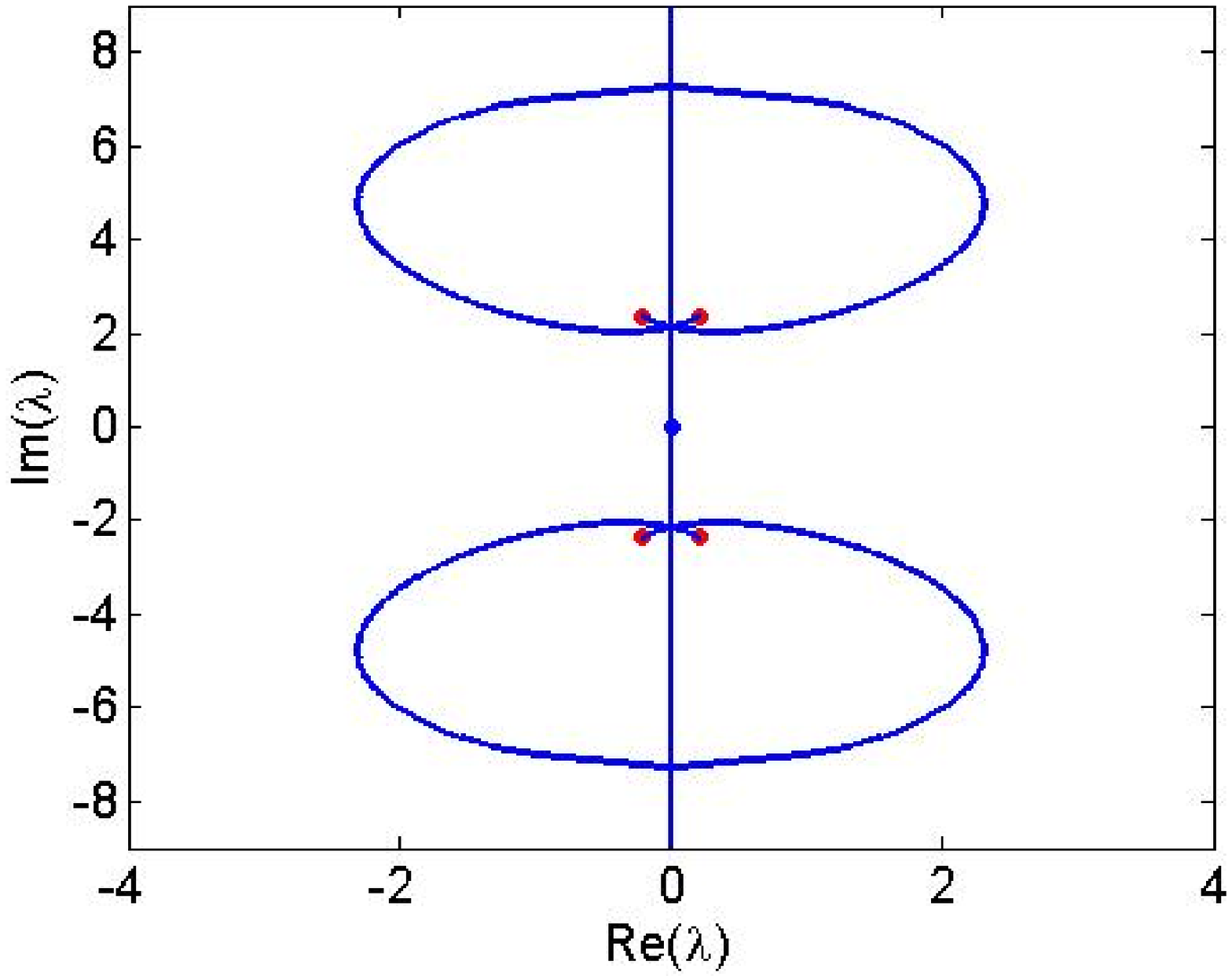}} \hfil
\subfigure[$\beta=\frac{\sqrt{2}}{2}$]{%
\includegraphics[height=50mm,width=65mm]{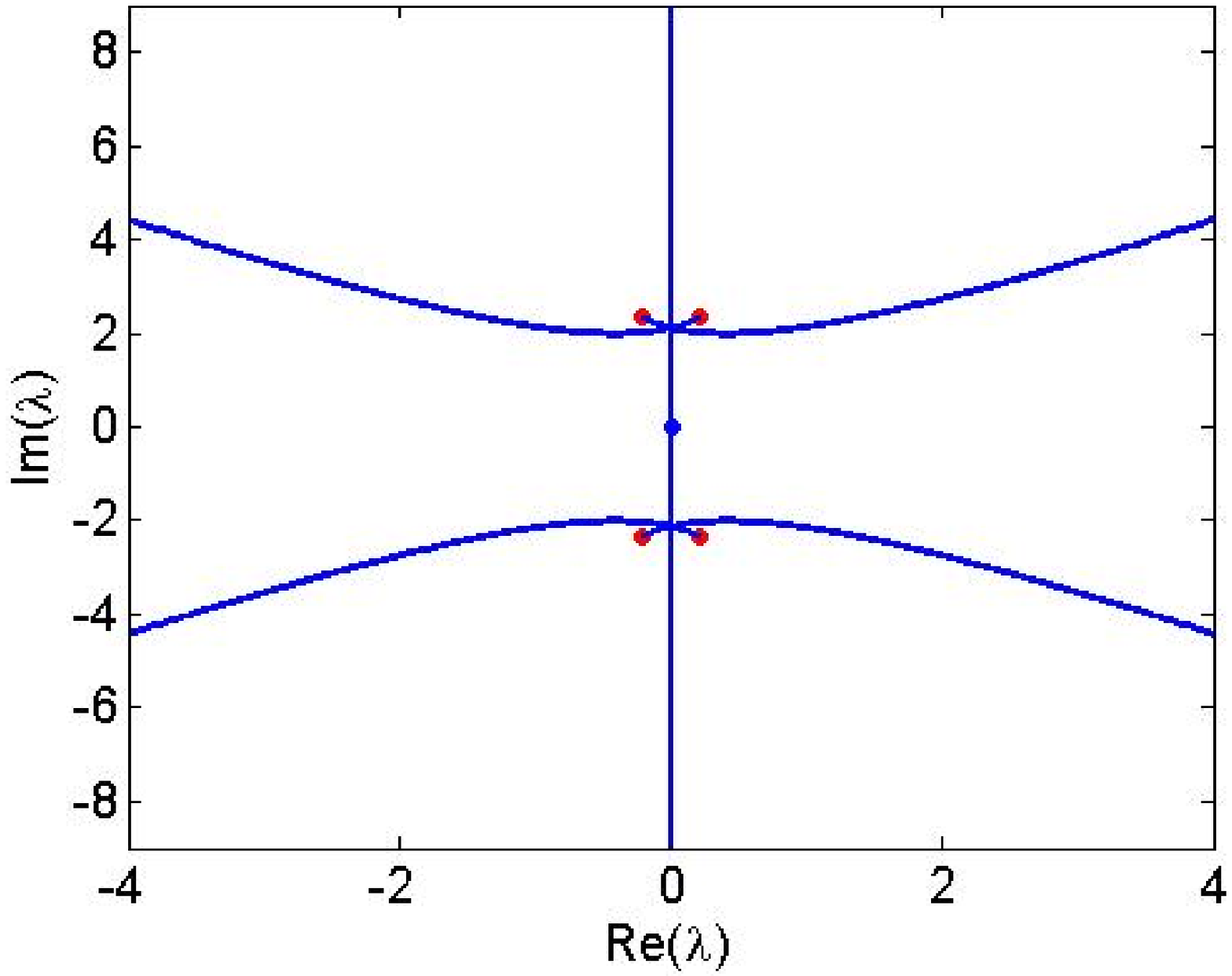}}
\caption{(color online): The red points represent the branch points for spectral curve $\lambda$.}
\label{con2}
\end{figure}

\begin{figure}[tbh]
\centering
\subfigure[$\beta=\frac{\sqrt{2}}{2}+\frac{1}{100}$]{%
\includegraphics[height=50mm,width=65mm]{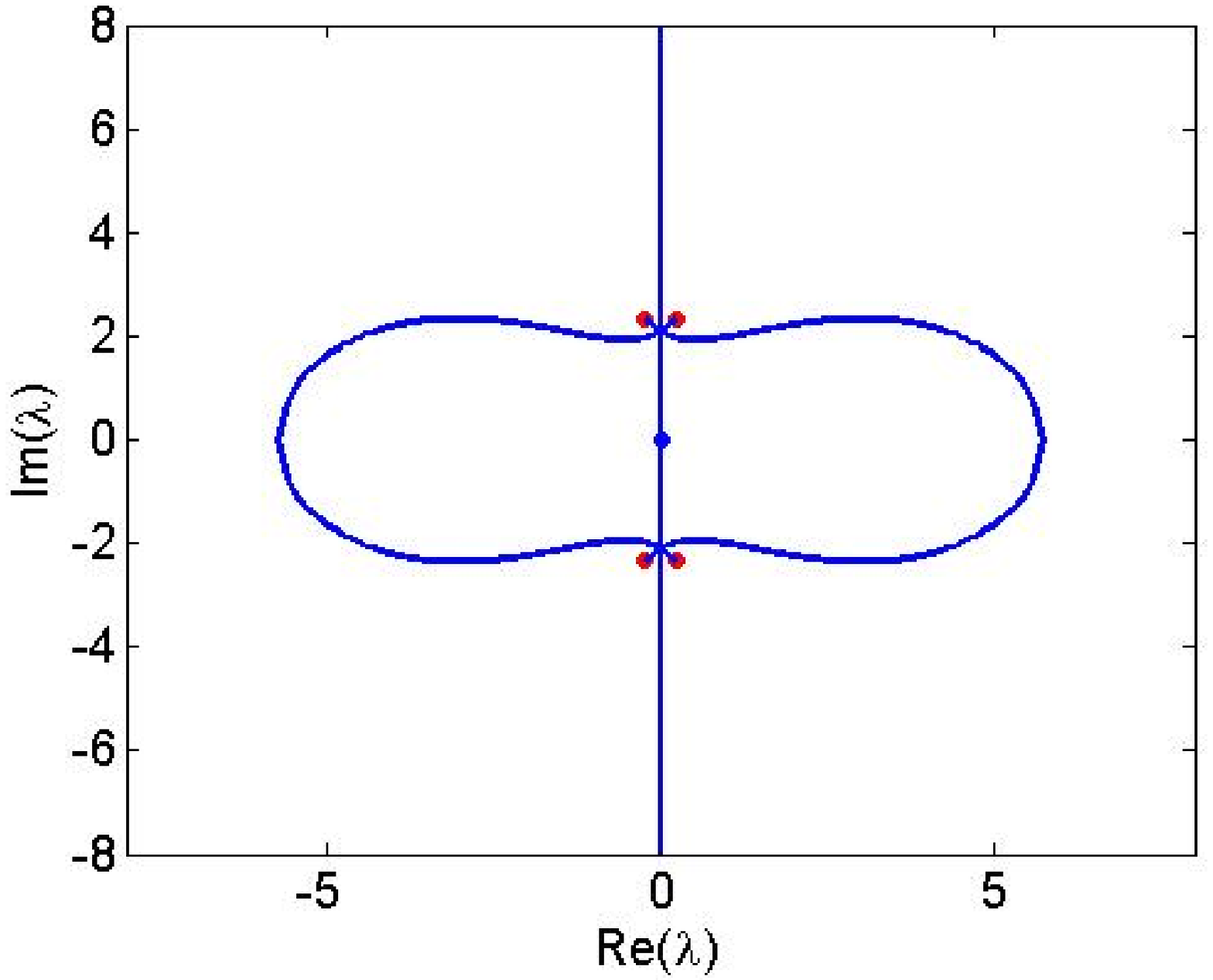}} \hfil
\subfigure[$\beta=2$]{%
\includegraphics[height=50mm,width=65mm]{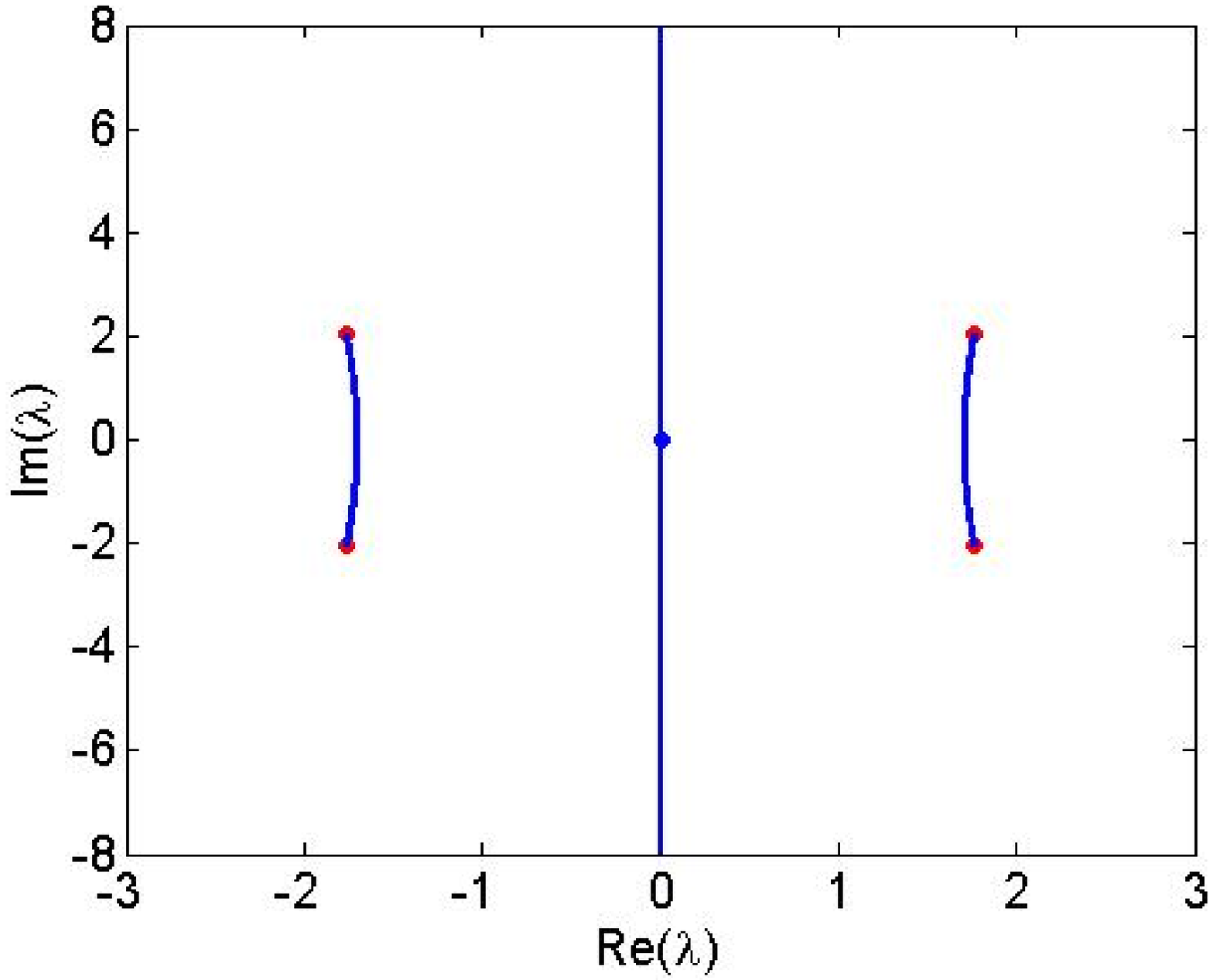}}
\caption{(color online): The red points represent the branch points for spectral curve $\lambda$.}
\label{con3}
\end{figure}

Through the above conditions, we can determine all spectral points $\lambda$ which satisfy that there are at least two roots of $\chi$ possessing the same imaginary part. On these points, we can obtain temporal homoclinic orbit solution. In what following, we illustrate how to determine the spectral points $\lambda$. Firstly, we can determine all the complex points $\chi_{1,\pm}$, $\chi_{2,\pm}$ through equations \eqref{equation1}. Then we insert them into formula
$\lambda=\chi-\frac{2\chi}{\chi^2-\beta^2}.$ For instance, we can plot the figure
with different parameter $\beta$ to show the variation of contour for $\lambda$. If we choose the parameter very small such as $\beta=\frac{1}{4}$, through above steps we can obtain the contour which are consisted of two rays and one segment (Fig. \ref{con1} (a)). Increase the parameter $\beta=\frac{1}{2}$, two rays and one segment merge into the image axis  (Fig. \ref{con1} (b)). When the parameter $\beta>\frac{1}{2}$ such as $\beta=\frac{7}{10}$, there appear four arcs (Fig. \ref{con2} (a)). Go on increasing the parameter to $\beta=\frac{\sqrt{2}}{2}$, in this case four arcs extends to the infinite (Fig. \ref{con2} (b)). If the parameter $\beta>\frac{\sqrt{2}}{2}$ such as $\beta=\frac{\sqrt{2}}{2}+\frac{1}{100}$, then we can obtain the contour with the terminal points of four arcs located on the real axis (Fig. \ref{con3} (a)). If the parameter $\beta$ continues to increase  $\beta=2$, then four arcs never interact each other (Fig. \ref{con3} (b)).

Nextly, we discuss them in three cases ($0<\beta<1/2$,  $1/2\leq \beta \leq \sqrt{2}/2$, and $\beta>\sqrt{2}/2$) to clarify the AB pattern in the two components. The AB pattern can be characterized by the structure of a unit period of AB.

\begin{figure}[tbh]
\centering
\includegraphics[height=60mm,width=150mm]{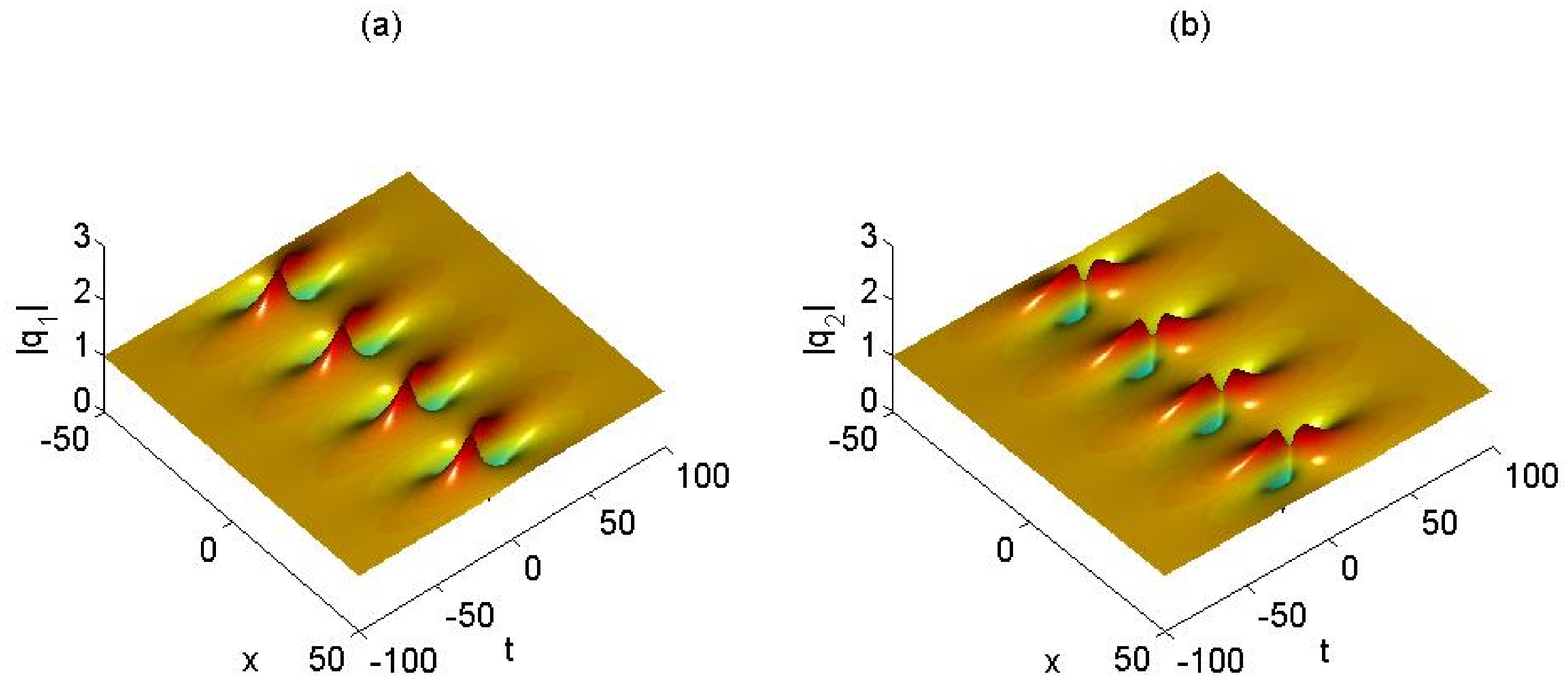} \hfil
\caption{(color online): The 3-D plot for $|q_1|$ and $|q_2|$.}
\label{fig1}
\end{figure}

\begin{figure}[tbh]
\centering
\includegraphics[height=60mm,width=150mm]{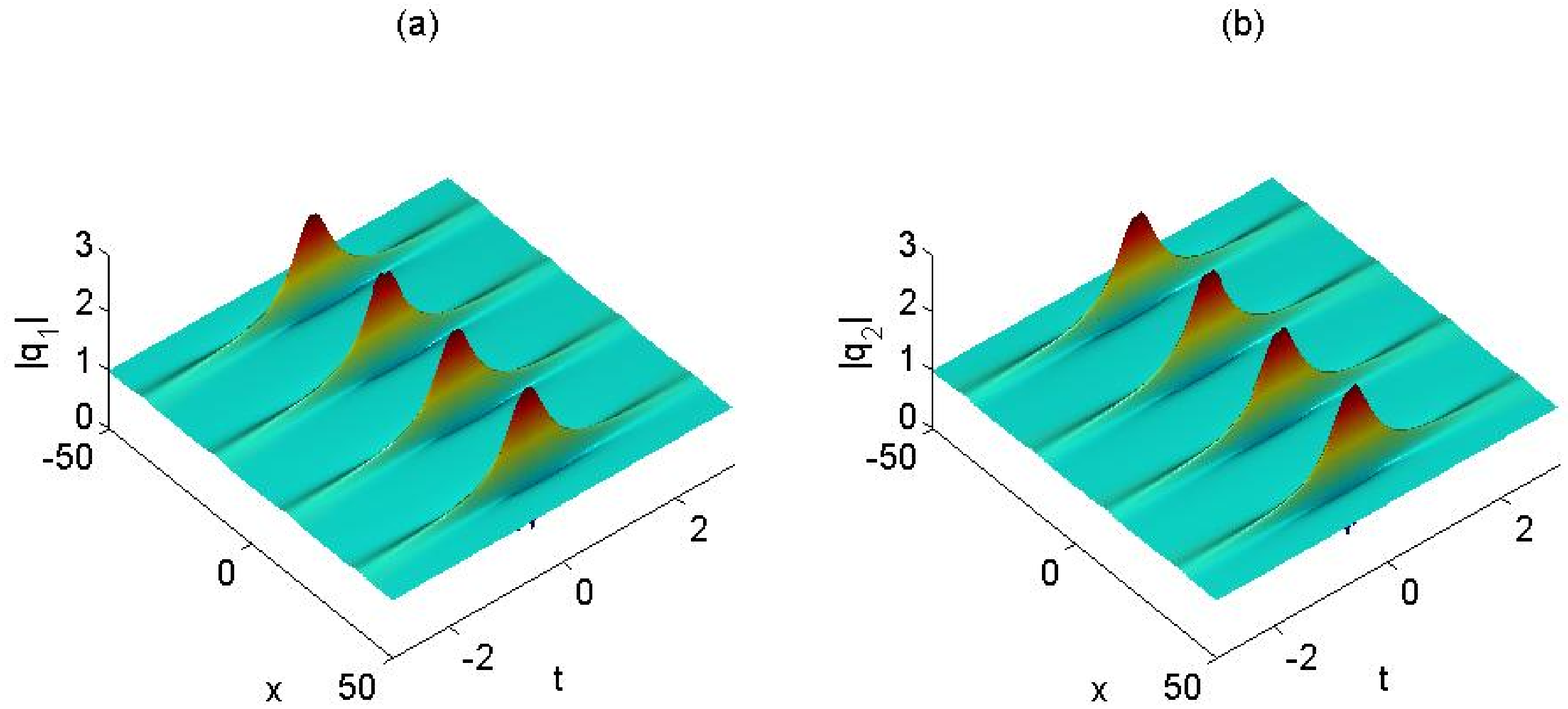} \hfil
\caption{(color online): The 3-D plot for $|q_1|$ and $|q_2|$.}
\label{fig2}
\end{figure}
\begin{figure}[tbh]
\centering
\includegraphics[height=60mm,width=150mm]{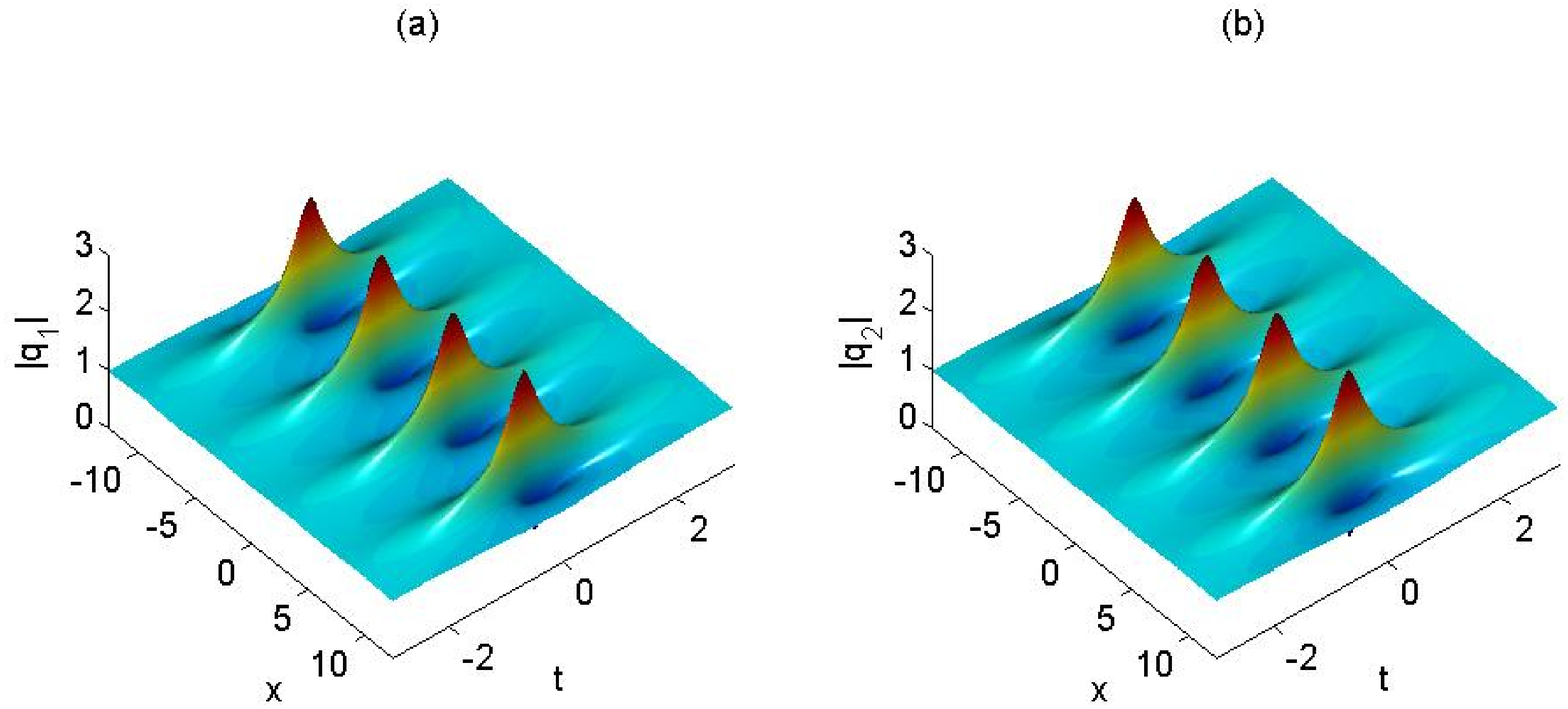} \hfil
\caption{(color online): The 3-D plot for $|q_1|$ and $|q_2|$.}
\label{fig3}
\end{figure}

\textbf{Case1: $0<\beta<1/2$}

Through above analysis, we can conclude that:
\begin{itemize}
  \item if $0<|\alpha|<2\beta$, there are two pairs of conjugated complex roots for equation \eqref{chara3}. The corresponding spectral parameter $\lambda$ locates on the image axis.
  \item if $2\beta\leq|\alpha|<2\sqrt{2+\beta^2}$, there exists a pair of conjugated complex root for equation \eqref{chara3}. The corresponding spectral parameter $\lambda$ also locates on the image axis.
\end{itemize}

In particular, we choose $\beta=\frac{1}{4}$ to discuss the AB pattern dynamics. As we can see from the Fig.  \ref{con1}(a), there are two rays extended to $\pm {\rm i}\infty$ and one line segment along the imaginary axis. And there are two gaps between the two rays and the line segment. This means that the spectral point locates on the contours for which AB solutions can be constructed. The two surplus rays are absent for scalar NLSE, therefore the two surplus rays are induced by cross-phase modulation effects. In the following, we would like to give the classification of different AB solution, according to spectral parameter locations on the contours.

Firstly, if we choose $\alpha=\frac{1}{4}<2\beta=\frac{1}{2}$, then there are two types of AB. i) The parameters can be solved exactly $\lambda\approx4.768{\rm i}$ (locates on the ray) , $\chi_1\approx0.125+0.232{\rm i}$, $\chi_2\approx-0.125+0.232{\rm i}$. Inserting them in to expressions, then we can obtain the dynamics of AB (Fig. \ref{fig1}). It is seen that
the structures of ABs are obviously distinctive in $q_1$ and $q_2$ components. ii) The other parameters can be solved $\lambda\approx2.772{\rm i}$ (locates on the line segment) , $\chi_1\approx0.125+1.338{\rm i}$, $\chi_2\approx-0.125+1.338{\rm i}$. Plugging them into formula \eqref{ele}, other dynamics can be obtained (Fig. \ref{fig2}). It is seen that
the structures of ABs in the two components are similar. Comparing with figure \ref{fig1} and figure \ref{fig2}, we infer that there
are complete different types of dynamical behavior for which spectral parameter locates on the ray and line segment. It should be pointed that these two different types of ABs can be superposed by above solitonic formula \eqref{gene-soliton}. But we omit them here.

Secondly, if we choose $\alpha=1>2\beta=\frac{1}{2}$, it follows that $\lambda\approx2.61 {\rm i}$ (locates on the line segment),  $\chi_1\approx0.5+1.261{\rm i}$, $\chi_2\approx-0.5+1.261{\rm i}$. Substituting these parameters into formula \eqref{ele}, we can obtain the figure (Fig. \ref{fig3}). It is seen that ABs in two components possesses the similar structures.

In summary, the location of spectral parameter $\lambda$ on the contour determines the AB dynamics in the two components. If the spectral parameter $\lambda$ locates on the line segment, we can obtain the ABs with the similar structures in two components. If the spectral parameter $\lambda$ locates on the ray extend to ${\rm i}\infty$,
the ABs possess different structures.

\textbf{Case 2: $1/2\leq \beta \leq \sqrt{2}/2$}

Similar as above case, we can conclude that:
\begin{itemize}
  \item if $0<|\alpha|<\sqrt{4-\beta^{-2}}$, there are two pairs of conjugated complex roots for equation \eqref{chara3}. The corresponding spectral parameter $\lambda$ locates on the arcs (see Fig. \ref{con2}a).
  \item if $\sqrt{4-\beta^{-2}}\leq|\alpha|<2\beta$, there are two pairs of conjugated complex roots for equation \eqref{chara3}. The corresponding spectral parameter $\lambda$ locates on the image axis.
  \item if $2\beta\leq|\alpha|<2\sqrt{2+\beta^2}$, there is a pair of conjugated complex root for equation \eqref{chara3}. The corresponding spectral parameter $\lambda$ locates on the image axis.
\end{itemize}

As an example, we choose $\beta=\frac{7}{10}$ to discuss AB dynamics.  The corresponding contour plotting for spectral parameter is shown in figure \ref{con2}(a).
It is shown that the contour is composed of the whole imaginary axis and two interacting arcs.

\begin{figure}[tbh]
\centering
\includegraphics[height=60mm,width=150mm]{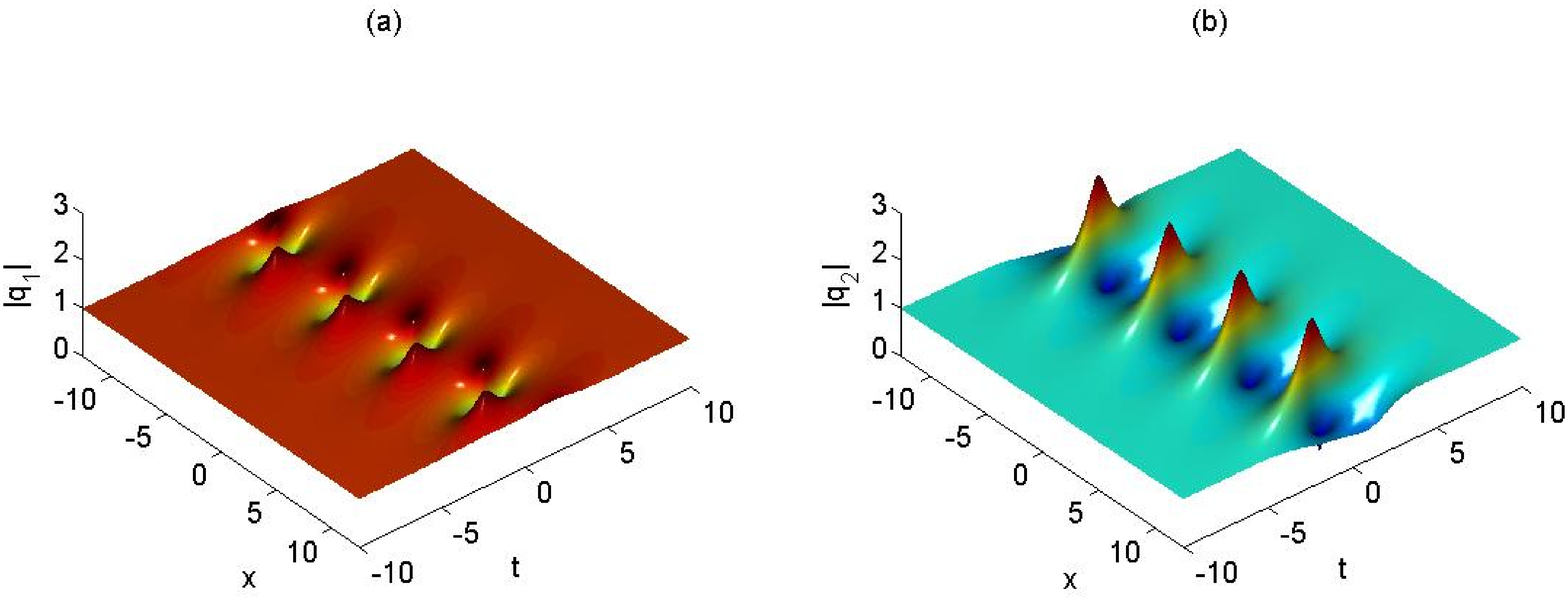} \hfil
\caption{(color online): The 3-D plot for $|q_1|$ and $|q_2|$.}
\label{fig4}
\end{figure}

Firstly $\alpha=1<\sqrt{4-\beta^{-2}}$ then there are two different types. In this case, we can obtain that $\lambda\approx-0.0134+2.134{\rm i}.$ The spectral point is located on the arc. It follows that $\chi_1\approx 0.986+0.705{\rm i}$, $\chi_2\approx-0.014+0.705{\rm i}$. Inserting these parameters into formula \eqref{ele}, we can plot the figure (Fig. \ref{fig4}). It is seen that ABs in the two component possess different structures. The other case $\lambda\approx0.0134+2.134{\rm i}$ is similar with above case. Thus we omit it. Similarly, we can iterate them by the solitonic formula \eqref{gene-soliton}.

\begin{figure}[tbh]
\centering
\includegraphics[height=60mm,width=150mm]{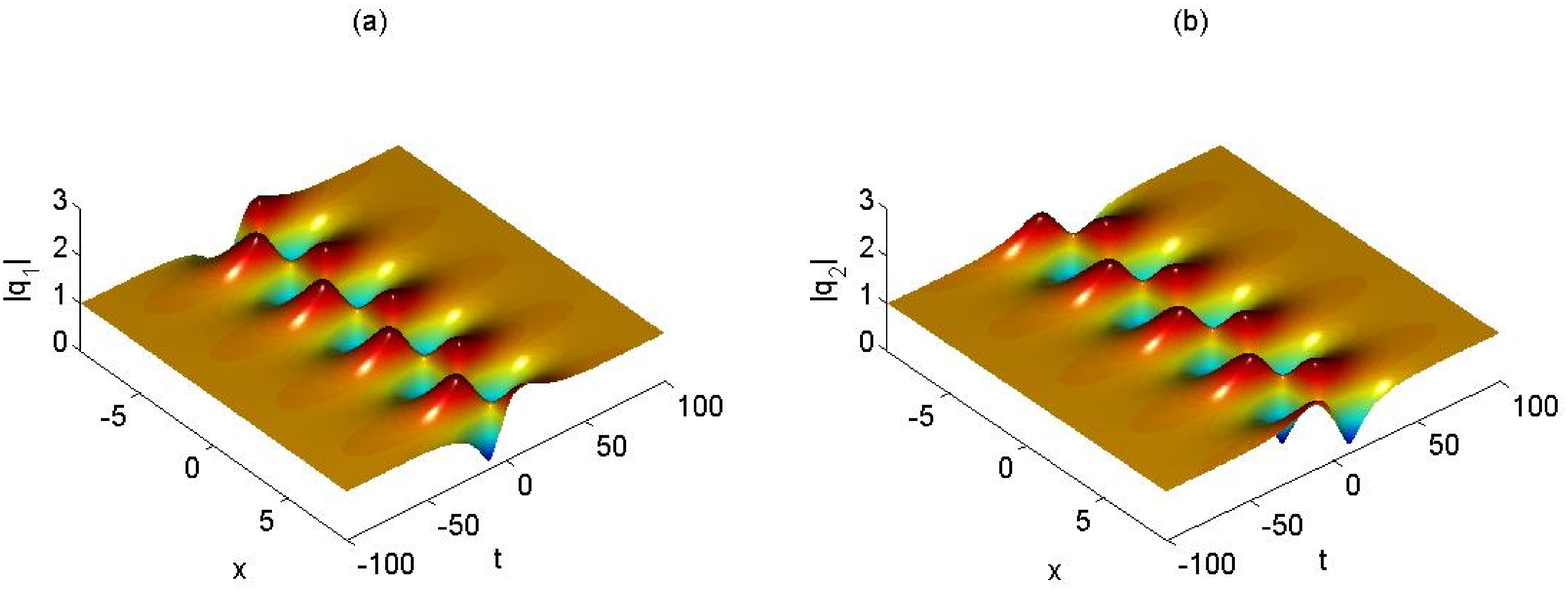} \hfil
\caption{(color online): The 3-D plot for $|q_1|$ and $|q_2|$.}
\label{fig5}
\end{figure}

\begin{figure}[tbh]
\centering
\includegraphics[height=60mm,width=150mm]{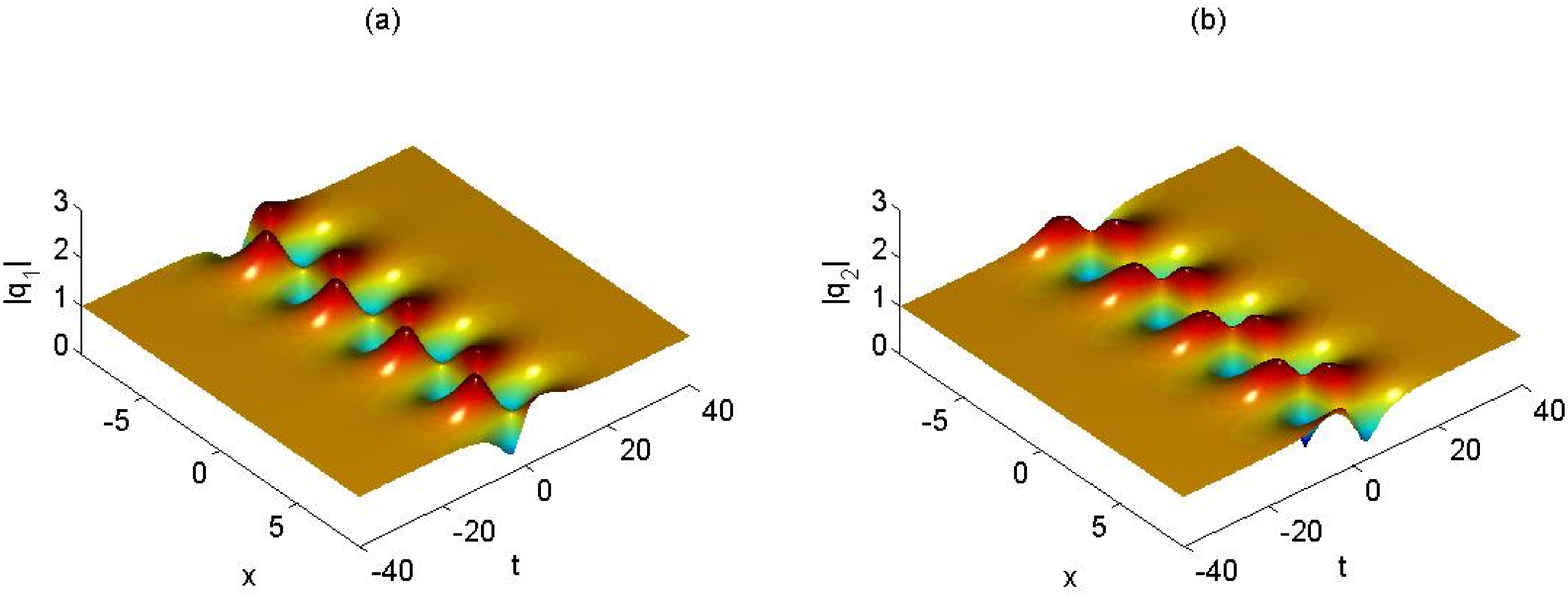} \hfil
\caption{(color online): The 3-D plot for $|q_1|$ and $|q_2|$.}
\label{fig6}
\end{figure}

Secondly, we choose $\sqrt{4-\beta^{-2}}<\alpha=\frac{14}{10}-10^{-4}<2\beta$, we can also obtain two kinds of ABs. i) $\lambda\approx16.192{\rm i}$ (locates on imaginary axis), $\chi_1\approx
0.69995+0.062{\rm i}$, $\chi_2\approx-0.69995+0.062{\rm i}$. Inserting them into formula \eqref{ele}, we can exhibit the dynamics (Fig. \ref{fig5}). In this case, we can see that ABs in the two components possess similar structure. ii) $\lambda\approx5.536{\rm i}$ (locates on imaginary axis), $\chi_1\approx.69995+.19{\rm i}$,
$\chi_2\approx-.69995+.19{\rm i}$. Inserting them into formula \eqref{ele}, we can exhibit the dynamics (Fig. \ref{fig6}). In this case, we can see that ABs in the two components possess similar structure.

\begin{figure}[tbh]
\centering
\includegraphics[height=60mm,width=150mm]{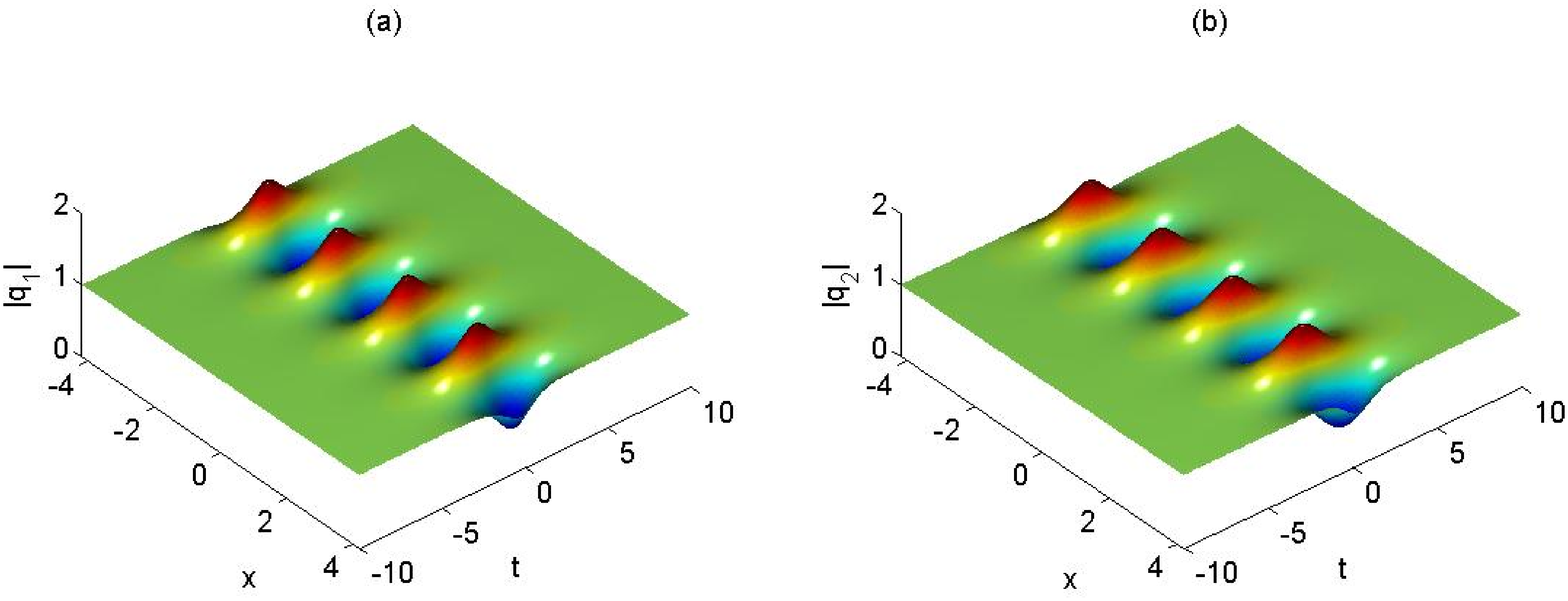} \hfil
\caption{(color online): The 3-D plot for $|q_1|$ and $|q_2|$.}
\label{fig7}
\end{figure}

Thirdly, we choose $\alpha=3>2\beta$, there merely exists a kind of AB.
The parameters can obtained as following: $\lambda\approx0.864{\rm i}$ (locates on imaginary axis), $\chi_1\approx1.5+0.343{\rm i}$,
$\chi_2\approx-1.5+.343{\rm i}$ (Fig. \ref{fig7}). It is seen that ABs in the two component possess similar structures.

In summary, if the spectral parameter $\lambda$ locates on imaginary axis, we can obtain the ABs with similar structures in the two components. If the spectral parameter $\lambda$ locates on the two arcs, the ABs possess different structures.

\textbf{Case 3: $\beta >\sqrt{2}/2$}

Similar as above cases, we can conclude that:
\begin{itemize}
  \item if $0<|\alpha|<\sqrt{4-\beta^{-2}}$, there are two pairs of conjugated complex roots for equation \eqref{chara3}. The corresponding spectral parameter $\lambda$ locates on the arcs (see Fig. \ref{con3}).
  \item if $2\beta\leq|\alpha|<2\sqrt{2+\beta^2}$, there is a pair of conjugated complex root for equation \eqref{chara3}. The corresponding spectral parameter $\lambda$ locates on the image axis.
\end{itemize}
Since this case is simpler than the \textbf{case 2}, we ignore the detailed discussion for this case.

\subsection{A special Akhmediev breather with different patterns }

In this subsection, we looking for the three roots with the same image part
$\chi$, $\chi+\alpha$, $\chi+2\alpha$. Here we merely give the existence condition for particular case
$a_1=a_2=1$, $\beta=\frac{1}{2}\sqrt{1+\alpha^2}$, $|\alpha|\in (0,\sqrt{3})$.

\begin{figure}[tbh]
\centering
\includegraphics[height=60mm,width=150mm]{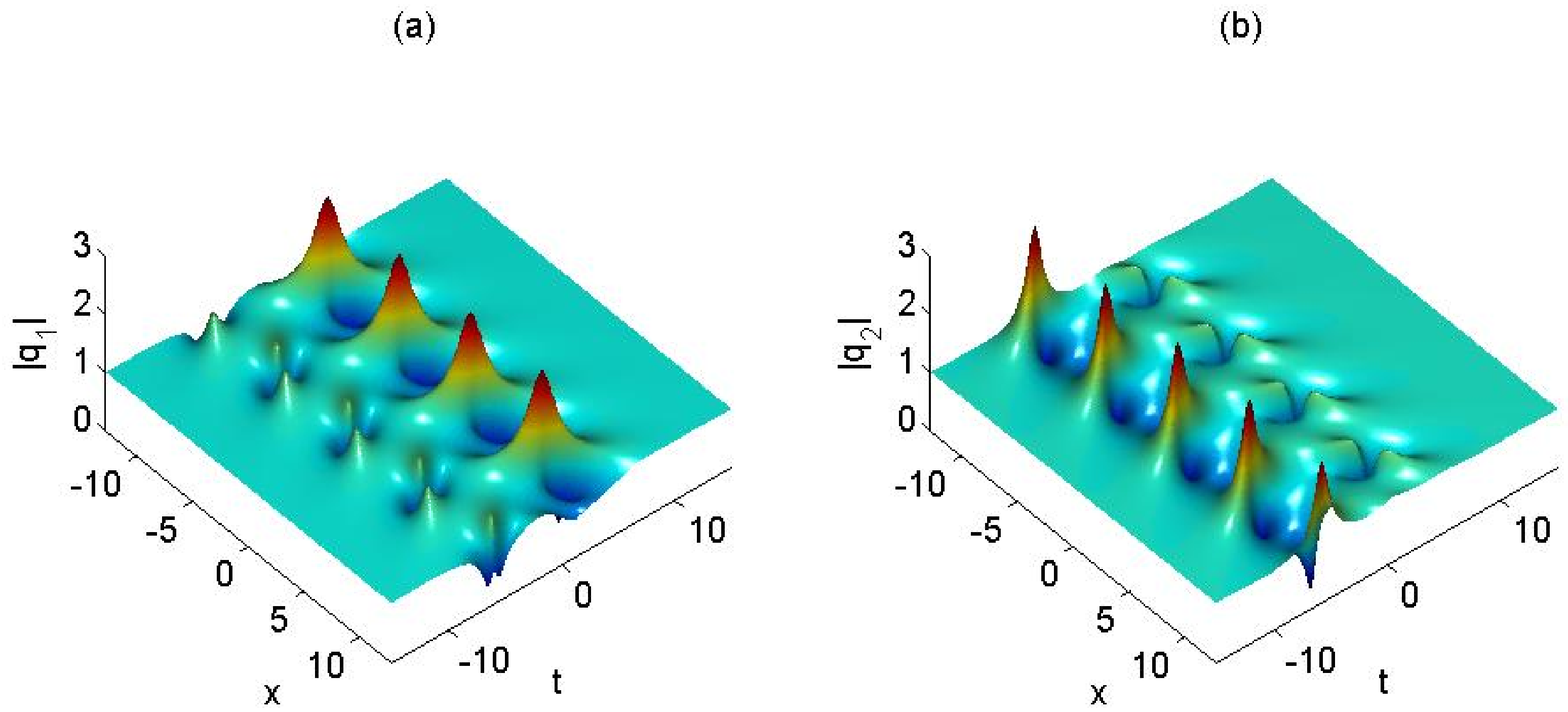} \hfil
\caption{(color online): The 3-D plot for $|q_1|$ and $|q_2|$.}
\label{fig8}
\end{figure}

\begin{figure}[tbh]
\centering
\includegraphics[height=60mm,width=150mm]{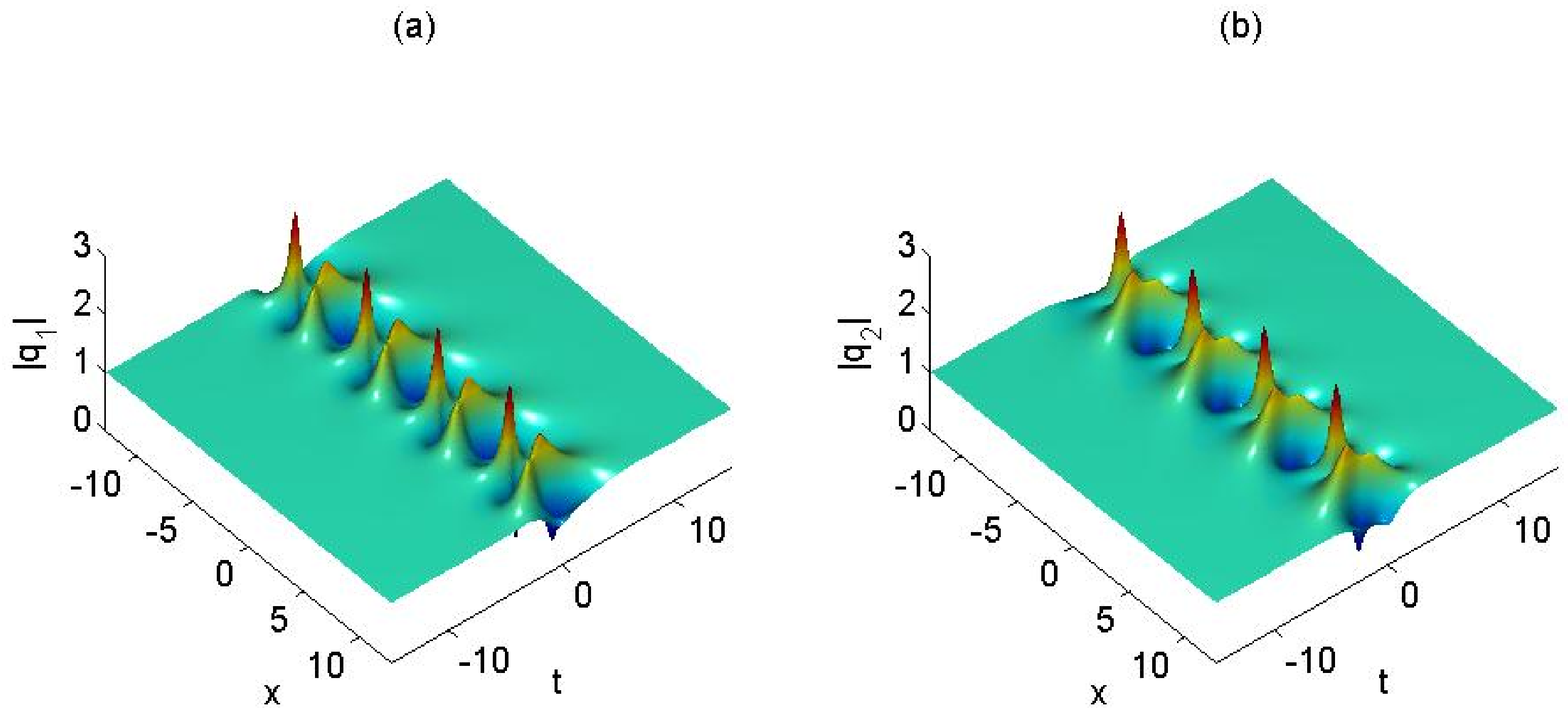} \hfil
\caption{(color online): The 3-D plot for $|q_1|$ and $|q_2|$.}
\label{fig9}
\end{figure}

\begin{figure}[tbh]
\centering
\includegraphics[height=60mm,width=150mm]{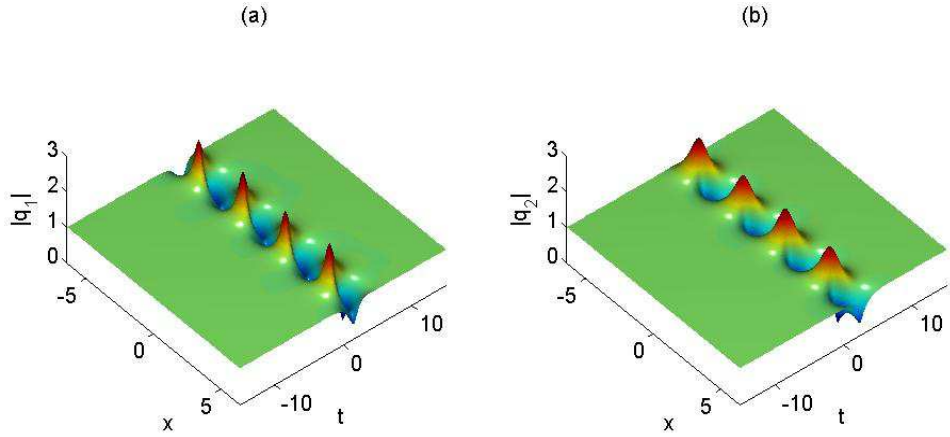} \hfil
\caption{(color online): The 3-D plot for $|q_1|$ and $|q_2|$.}
\label{fig10}
\end{figure}

Especially, we need to consider the special case for which there are three different branches of curve cross at the point of image axis, $a_1=a_2=1$, $\beta=\frac{7}{10}$, $\alpha=\frac{2}{5}\sqrt{6}$, $\lambda=\frac{3}{10}\sqrt{51}{\rm i}$, $\chi_0=-\frac{2}{5}\sqrt{6}+\frac{\sqrt{51}}{10}{\rm i},\chi_1=\frac{\sqrt{51}}{10}{\rm i}, \chi_2=\frac{2}{5}\sqrt{6}+\frac{\sqrt{51}}{10}{\rm i}$. In this case, we use the following formula to describe the dynamics of a superposition of ABs:
\begin{equation*}
  q_i=a_i\frac{M_i}{M}{\rm e}^{\theta_i},\,\, i=1,2,
\end{equation*}
where
\begin{equation*}
  \begin{split}
     M=&\sum_{l=0,s=0}^{2,2}\frac{\exp(\omega_{l}^*+\omega_{s})}{\chi_{s}-\chi_{l}^*},  \\
     M_i=&\sum_{l=0,s=0}^{2,2}\frac{\chi_{s}^*+b_i}{\chi_{l}+b_i}\frac{\exp(\omega_{l}^*+\omega_{s})}{\chi_{s}^*-\chi_{l}},\,\, \omega_0=0,\\
     \omega_{k}=&{\rm i}k\alpha\left( (x-x_k)+\frac{1}{2}(\chi_{k}+\chi_{0})(t-t_k)\right),\,\, x_k,\,\,t_k\in \mathbb{R},\,\, k=1,2.
  \end{split}
\end{equation*}
It is very interesting that there are two complete different dynamics behavior by choosing the parameter like the following:

\begin{description}
  \item[a)] If we choose the parameters $t_1=-t_2=2$, $x_1=x_2=0$, then we obtain a double- breather (Fig. \ref{fig8}), which admits two different AB excitation patterns.
  \item[b)] If we choose the parameters $t_1=t_2=0$, $x_1=x_2=0$, this is an immediate state (Fig. \ref{fig9}), which is an overlapping superposition of the above two ABs.
  \item[c)] If we choose the parameters $t_1=-t_2=-2$, $x_1=x_2=0$, we obtain one breather solution with double period (Fig. \ref{fig10}) since the terms $\omega_0$ and $\omega_2$ play an important role in determining the dynamics.
\end{description}

The above two breathers with an identical spectral parameter can be seen as a special double-breather. Furthermore, we can iterate multi-breather with different spectral parameters, which can admit many different superposition patterns. On the other hand, we can iterate high-order breather with an identical spectral parameter. But which can not be explained by the MI analysis. As for the branch points which correspond to the rogue wave solution \cite{Yang,Ling,gaillard,clarkson,he0,Chow}, we would like to discuss them in the following.

\section{The general rogue wave solutions for VNLSE}\label{sec4}
If the above algebraic equation \eqref{cubic} possesses a multiple root,  RW  solutions can be obtained. The RW solution for VNLSE is different from the case for scalar NLSE for which spectral parameter must be fixed to be one value for deriving high-order RWs. The RW in vector case can admit more freedom for the spectral parameter. This provides possibilities to obtain superposition of RWs with different structures.  In other words, the spectral parameter
$\lambda_i$ must satisfy the following algebraic equation (discriminant equation for equation \eqref{cubic}):
\begin{equation}\label{discrim}
\prod_{i=1}^{N}(\chi+b_i)^2\left(1+\sum_{i=1}^{N}\frac{a_i^2}{(\chi+b_i)^2}\right)=0.
\end{equation}
It is readily to see that the discriminant equation \eqref{discrim} possesses $N$ pairs of complex conjugated roots if the parameters $b_i$s are not equal to each other (multiple roots are counted by its multiplicity). This makes the spectral parameter admit much more values to obtain RW solutions. Thus we can denote them as $\chi_1$, $\chi_1^*$, $\chi_2$, $\chi_2^*$, $\cdots,$ $\chi_p$, $\chi_p^*$, where
$\mathrm{Im}(\chi_j)>0,$ $j=1,2,\cdots,p,$ the multiplicity for each roots $\chi_i$, $\chi_i^*$ is $\kappa_i$ and
$\sum_{i=1}^{p}\kappa_i=N.$

Here we develop the formal series to tackle with high-order rational solutions for VNLSE. Fixed the parameters
$a_i$ and $b_i$, then
we have the following lemmas:
\begin{lem}\label{lem1}
The formal series
\begin{equation}\label{asymp}
\begin{split}
  \widehat{\lambda_j}&=\lambda_j^{[0]}+{\alpha_j}\epsilon_j^{\kappa_j+1},\,\,   \\
  \widehat{\chi_j}&=\sum_{m=0}^{\infty}\chi_j^{[m]}\epsilon_j^m,\,\, \chi_j^{[0]}=\chi_j,
\end{split}
\end{equation}
satisfy the $N+1$-th order algebraic equation \eqref{cubic}.
The $\epsilon_j$ is a small complex parameter.
The parameters $\chi_j^{[1]}=1$ and $\chi_j^{[m\geq 2]}$ can be determined recursively
\begin{equation*}
  \begin{split}
      \lambda_j^{[0]}=&\left(\chi_j^{[0]}-\sum_{i=1}^{N}\frac{a_i^2}{\chi_j^{[0]}+b_i}\right),  \\
      \alpha_j=&\sum_{i=1}^{N}a_i^2\left(-\frac{1}{\chi_j^{[0]}+b_i}\right)^{\kappa_j+2},  \\
      \chi_j^{[m-\kappa_j]}=&\frac{1}{(\kappa_j+1){\displaystyle \sum_{i=1}^{N}a_i^2\left(\frac{-1}{\chi_j^{[0]}+b_i}\right)^{\kappa_j+2}}}  \sum_{i=1}^{N}\left[\frac{a_i^2}{\chi_j^{[0]}+b_i}\sum_{\Lambda_1(m,\kappa_j)=m,\,\,
      l_{\alpha}\geq 0}^{
      {\Lambda(m,\kappa_j)\geq \kappa_j+1}}\right.\\
      &\left.\Lambda(m,\kappa_j)!\left(\frac{-1}{\chi_j^{[0]}+b_i}\right)^{\Lambda(m,\kappa_j)}\prod_{\alpha=1}^{m-\kappa_j-1}\frac{(\chi_j^{[\alpha]})^{l_{\alpha}}}{l_{\alpha}!}\right],\,\, m\geq \kappa_j-2,
  \end{split}
\end{equation*}
where $\Lambda(m,\kappa_j)={\displaystyle\sum_{\alpha=1}^{m-\kappa_j-1}l_{\alpha}}$, $\Lambda_1(m,\kappa_j)={\displaystyle\sum_{\alpha=1}^{m-\kappa_j-1}\alpha l_{\alpha}}.$
\end{lem}
\textbf{Proof:}
We can prove this lemma by the following steps:

\textbf{Step 1:} If $\chi_j^{[0]}$ possesses $\kappa_j+1$ multiplicity for the algebraic equation \eqref{cubic}, then the following identities are
verified:
\begin{equation}\label{identi}
  \begin{split}
     1+\sum_{i=1}^{N}\frac{a_i^2}{(\chi_j^{[0]}+b_i)^2} &=0,  \\
     \sum_{i=1}^{N}\frac{a_i^2}{(\chi_j^{[0]}+b_i)^3}=&0, \\
     \vdots&\\
      \sum_{i=1}^{N}\frac{a_i^2}{(\chi_j^{[0]}+b_i)^{\kappa_j+1}}=&0.
  \end{split}
\end{equation}

\textbf{Step 2:} We have the following expansions:
\begin{equation*}
  \begin{split}
      \frac{1}{(\chi_j^{[0]}+b_i)+{\displaystyle\sum_{m=1}^{\infty}\chi_j^{[m]}\epsilon_j^{m}}}=&\frac{1}{\chi_j^{[0]}+b_i}\frac{1}{
      1+{\displaystyle\sum_{m=1}^{\infty}\frac{\chi_j^{[m]}}{\chi_j^{[0]}+b_i}\epsilon_j^{m}}}  \\
      =&\frac{1}{\chi_j^{[0]}+b_i}\sum_{k=0}^{\infty}\left(-\sum_{m=1}^{\infty}\frac{\chi_j^{[m]}}{
      \chi_j^{[0]}+b_i}\epsilon_j^{m}\right)^{k}\\
      =&\sum_{m=0}^{\infty}\widehat{\chi}_{j,i}^{[m]}\epsilon_j^{m},
  \end{split}
\end{equation*}
where
\begin{equation*}
  \widehat{\chi}_{j,i}^{[m]}=\frac{1}{\chi_j^{[0]}+b_i}\left[\sum_{\sum_{\alpha=1}^{m}\alpha l_{\alpha}=m}^{l_{\alpha}\geq 0}\left(\sum_{\alpha=1}^{m}l_{\alpha}\right)!\prod_{\alpha=1}^{m}\left(\frac{1}{
      l_{\alpha}!}\left(-\frac{\chi_j^{[\alpha]}}{\chi_j^{[0]}+b_i}\right)^{l_{\alpha}}\right)\right].
\end{equation*}

\textbf{Step 3:} It follows that the characteristic equation \eqref{cubic} can be expanded as the following:
\begin{equation*}
\begin{split}
  &\chi_j^{[0]}+\sum_{m=1}^{\infty}\chi_j^{[m]}\epsilon_j^{m}-(\lambda_j^{[0]}+\alpha_j\epsilon_j^{\kappa_j+1})
  \\
    &-\sum_{i=1}^{N}\frac{a_i^2}{\chi_j^{[0]}+b_i}\sum_{m=0}^{\infty}\left[\sum_{\sum_{\alpha=1}^{m}\alpha l_{\alpha}=m}^{l_{\alpha}\geq 0}\left(\sum_{\alpha=1}^{m}l_{\alpha}\right)!\prod_{\alpha=1}^{m}\left(\frac{1}{
      l_{\alpha}!}\left(-\frac{\chi_j^{[\alpha]}}{\chi_j^{[0]}+b_i}\right)^{l_{\alpha}}\right)\right]\epsilon_j^{m}=0.
\end{split}
\end{equation*}
Comparing the coefficients $\epsilon_j$ of above equation together with equations \eqref{identi}, we arrive at
\begin{equation}\label{coeffk}
  \begin{split}
    \epsilon_j^{\kappa_j+1}:\,\,\,&-\alpha_j+\sum_{i=1}^{N}a_i^2\left(-\frac{1}{\chi_j^{[0]}+b_i}\right)^{\kappa_j+2}
    (\chi_j^{[1]})^{\kappa_j+1}=0,
  \end{split}
\end{equation}
\begin{equation}\label{coeffk1}
  \epsilon_j^{m},\,(m\geq \kappa_j+2): \,\, \sum_{i=1}^{N}\frac{a_i^2}{\chi_j^{[0]}+b_i}\sum_{{\sum_{\alpha=1}^{m-\kappa_j}\alpha l_{\alpha}}=m,\,\,
      l_{\alpha}\geq 0}^{
      {\sum_{\alpha=1}^{m-\kappa_j}}l_{\alpha}\geq \kappa_j+1}\left(\sum_{\alpha=1}^{m-\kappa_j}l_{\alpha}\right)!\left(\frac{-1}{\chi_j^{[0]}+b_i}\right)^{\sum_{\alpha=1}
      ^{m-\kappa_j}l_{\alpha},}\prod_{\alpha=1}^{m-\kappa_j}\frac{(\chi_j^{[\alpha]})^{l_{\alpha}}}{l_{\alpha}!}=0.
\end{equation}
Through the equation \eqref{coeffk}, we set $\chi_j^{[1]}=1$, it follows that
\begin{equation*}
  \alpha_j=\sum_{i=1}^{N}a_i^2\left(-\frac{1}{\chi_j^{[0]}+b_i}\right)^{\kappa_j+2}.
\end{equation*}
Furthermore, with the aid of the equation \eqref{coeffk1},
it follows that
\begin{equation*}
  \begin{split}
    \chi_j^{[m-\kappa_j]}=&\frac{1}{(\kappa_j+1){\displaystyle \sum_{i=1}^{N}a_i^2\left(\frac{-1}{\chi_j^{[0]}+b_i}\right)^{\kappa_j+2}}}  \sum_{i=1}^{N}\left[\frac{a_i^2}{\chi_j^{[0]}+b_i}\sum_{{\sum_{\alpha=1}^{m-\kappa_j-1}\alpha l_{\alpha}}=m,\,\,
      l_{\alpha}\geq 0}^{
      {\sum_{\alpha=1}^{m-\kappa_j-1}}l_{\alpha}\geq \kappa_j+1}\right.\\
      &\left.\left(\sum_{\alpha=1}^{m-\kappa_j-1}l_{\alpha}\right)!\left(\frac{-1}{\chi_j^{[0]}+b_i}\right)^{\sum_{\alpha=1}
      ^{m-\kappa_j-1}l_{\alpha},}\prod_{\alpha=1}^{m-\kappa_j-1}\frac{(\chi_j^{[\alpha]})^{l_{\alpha}}}{l_{\alpha}!}\right],\,\, m\geq \kappa_j-2.
  \end{split}
\end{equation*}
With this procedure, we can find that the coefficients $\chi_j^{[m\geq 2]}$ can be determined by above relations recursively.
$\square$

Since $\widehat{\chi_i}$ can be expanded with the series of $\epsilon_i$,
it follows that
\begin{equation*}
   \widehat{\omega_i}={\rm i}\left[\widehat{\chi_i} x+\frac{1}{2}\widehat{\chi_i}^2t\right]+\sum_{k=0}^{\infty}\beta_i^{[k]}\epsilon_i^{k}=\sum_{k=0}^{\infty}X_i^{[k]}\epsilon_i^{k}
\end{equation*}
where
\begin{equation*}
  X_i^{[k]}={\rm i}\left(\chi_i^{[k]}x+\frac{1}{2}\sum_{j=0}^{k}\chi_i^{[j]}\chi_i^{[k-j]}t\right).
\end{equation*}
Particularly, we can know that the first three terms for $X_i^{[k]}$ are
\begin{equation*}
    \begin{split}
      X_i^{[1]}=&{\rm i} \left(x+\chi_it\right)+\beta_i^{[1]}, \\
      X_i^{[2]}=&{\rm i} \left[\chi_i^{[2]} x+\left(\chi_i^{[2]}\chi_i+\frac{1}{2}\right)t\right]+\beta_i^{[2]},\\
      X_i^{[3]}=&{\rm i} \left[\chi_i^{[3]} x+\left(\chi_i^{[3]}\chi_i+\chi_i^{[2]} \right) t
 \right]+\beta_i^{[3]}.
    \end{split}
\end{equation*}
Moreover, based on the elementary Schur polynomials we have an  expansion
\begin{lem}\label{lem2}
\begin{equation*}
  \exp\left(\sum_{k=1}^{\infty}X_i^{[k]}\epsilon_i^{k}\right)=\sum_{j=0}^{\infty}S_i^{[j]}\epsilon_i^{j}
\end{equation*}
where $S_i^{[j]}$ is
\begin{equation*}
  S_i^{[j]}=\sum_{{\sum_{k=0}^{m}kl_k=j}}\frac{(X_i^{[1]})^{l_1}(X_i^{[2]})^{l_2}\cdots (X_i^{[m]})^{l_m}}{l_1!l_2!\cdots l_m!}.
\end{equation*}
\end{lem}
Specially
\begin{equation*}
    \begin{split}
    S_i^{[0]}=&1,\,\, S_i^{[1]}=X_i^{[1]},\\
      S_i^{[2]}=&\frac{1}{2}(X_{i}^{[1]})^{2}+X_{i}^{[1]},\\
      S_i^{[3]}=&X_{i}^{[3]}+X_{i}^{[1]}X_{i}^{[2]}+\frac{1}{6}(X_{i}^{[1]})^{3}.
    \end{split}
\end{equation*}

On the other hand, we need the following expansion series
\begin{equation*}
  \begin{split}
      \frac{1}{\widehat{\chi_k}^*-\widehat{\chi_j}}=&\left[\chi_k^{[0]*}-\chi_j^{[0]}+\sum_{m=1}^{\infty}(\chi_k^{[m]*}
      \epsilon_k^{*m}-\chi_j^{[m]}
      \epsilon_j^{m})\right]^{-1}  \\
      =&\frac{1}{\chi_k^{[0]*}-\chi_j^{[0]}}\sum_{n=0}^{\infty}\left[-\frac{1}{\chi_k^{[0]*}-\chi_j^{[0]}}\sum_{m=1}^{\infty}(\chi_k^{[m]*}
      \epsilon_k^{*m}-\chi_j^{[m]}
      \epsilon_j^{m})\right]^{n}\\
      =&\sum_{m=0,n=0}^{\infty,\infty}E_{k,j}^{[m,n]}\epsilon_k^{*m}\epsilon_j^{n}
  \end{split}
\end{equation*}
where
\begin{equation*}
  \begin{split}
     E_{k,j}^{[m,n]}=&\frac{1}{\chi_k^{[0]*}-\chi_j^{[0]}}\sum_{L_m=m,H_n=n}\left(-
     \frac{1}{\chi_k^{[0]*}-\chi_j^{[0]}}\right)^{\Lambda_{m,n}} \Lambda_{m,n}!\prod_{\alpha=1}^{m}
     \left(\frac{(\chi_k^{[\alpha]})^{l_{\alpha}}}{l_{\alpha}!}\right)^* \prod_{\beta=1}^{n}
     \left(\frac{(-\chi_j^{[\beta]})^{l_{\beta}}}{l_{\beta}!}\right)\\
      L_m=&\sum_{\alpha=1}^{m}\alpha l_{\alpha},\,\, H_n=\sum_{\beta=1}^{n}\beta l_{\beta},\,\,
      \Lambda_{m,n}=\sum_{\alpha=1}^{m}l_{\alpha}+\sum_{\beta=1}^{n}l_{\beta},\,\, l_{\alpha},l_{\beta}\geq0.
  \end{split}
\end{equation*}
In a similar way, we can expand
\begin{equation*}
  \begin{split}
      \frac{1}{\widehat{\chi_k}^*-\widehat{\chi_j}}\frac{\widehat{\chi_k}^*+b_i}{\widehat{\chi_j}+b_i}=&\left(\sum_{m=0,n=0}^{\infty,\infty}E_{k,j}^{[m,n]}\epsilon_k^{*m}\epsilon_j^{n}\right)
      \left(\sum_{m=0,n=0}^{\infty,\infty}\chi_{k,i}^{[m]*}\widehat{\chi}_{j,i}^{[n]}\epsilon_k^{*m}\epsilon_j^{n}\right)\\
      =&\sum_{m=0,n=0}^{\infty,\infty}E_{k,j}^{[i;l,s]}\epsilon_k^{*m}\epsilon_j^{n}
  \end{split}
\end{equation*}
where
\begin{equation*}
  E_{k,j}^{[i;l,s]}=\sum_{l=0,s=0}^{m,n}E_{k,j}^{[l,s]}\chi_{k,i}^{[m-l]*}\widehat{\chi}_{j,i}^{[n-s]}.
\end{equation*}
Then the elements of matrix can be expanded as
\begin{equation*}
\begin{split}
      \frac{{\rm e}^{\omega_k^*+\omega_j}}{\widehat{\chi_k}^*-\widehat{\chi_j}}&=\left(\sum_{m=0,n=0}^{\infty,\infty}E_{k,j}^{[m,n]}\epsilon_k^{*m}\epsilon
  _j^{n}\right)\left(\sum_{m=0,n=0}^{\infty,\infty}S_k^{[m]*}S_j^{[n]}\epsilon_k^{*m}\epsilon_j^{n}\right) \\
  &=\sum_{m=0,n=0}^{\infty,\infty}H_{k,j}^{[m,n]}\epsilon_k^{*m}\epsilon_j^{n}
\end{split}
\end{equation*}
where
\begin{equation*}
  H_{k,j}^{[m,n]}=\sum_{l=0,s=0}^{m,n}E_{k,j}^{[l,s]}S_k^{[m-l]*}S_j^{[n-s]}.
\end{equation*}
In a similar way, we have
\begin{equation*}
\begin{split}
      \frac{{\rm e}^{\omega_k^*+\omega_j}}{\widehat{\chi_k}^*-\widehat{\chi_j}}\frac{\widehat{\chi_k}^*+b_i}{\widehat{\chi_j}+b_i}&=\left(\sum_{m=0,n=0}^{\infty,\infty}E_{k,j}^{[i;m,n]}\epsilon_k^{*m}\epsilon
  _j^{n}\right)\left(\sum_{m=0,n=0}^{\infty,\infty}S_k^{[m]*}S_j^{[n]}\epsilon_k^{*m}\epsilon_j^{n}\right) \\
  &=\sum_{m=0,n=0}^{\infty,\infty}H_{k,j}^{[i;m,n]}\epsilon_k^{*m}\epsilon_j^{n}
\end{split}
\end{equation*}
where
\begin{equation*}
  H_{k,j}^{[i;m,n]}=\sum_{l=0,s=0}^{m,n}E_{k,j}^{[i;l,s]}S_k^{[m-l]*}S_j^{[n-s]}.
\end{equation*}

Denote the vector solution
\begin{equation*}
    \Phi_{j}(\epsilon_j)=D_{N+1}\begin{bmatrix}
             \exp{\omega_{j}}
             ,\,
             \displaystyle{\frac{\exp{\omega_{j}}}{\widehat{\chi_{j}}+b_1}},\,
             \cdots, \,
             \displaystyle{\frac{\exp{\omega_{j}}}{\widehat{\chi_{j}}+b_N}}
           \end{bmatrix}^T,\,\, \omega_j={\rm i}\left[\widehat{\chi_j} x+\frac{1}{2}\widehat{\chi_j}^2t\right]+\beta_j^{[0]}.
\end{equation*}
We can deduce the symmetry relations for solutions $\Phi_j(\epsilon_j)$ to construct the other vector solutions. If $\Phi_j(\epsilon_j)$ is a solution for Lax pair
\eqref{Lax}, then $\Phi_j\left(\epsilon_j\exp{\left(\frac{2l_j\pi {\rm i}}{\kappa_j+1}\right)}\right)$ is also a solution for system
\eqref{Lax}, $0\leq l_j\leq \kappa_j+1$. We expand them as
\begin{equation*}
  \begin{split}
     \Phi_j(\epsilon_j)=&\sum_{m=0}^{\infty}\Phi_j^{[m]}\epsilon_j^{m},  \\
     \Phi_j\left({\rm e}^{\frac{2\pi {\rm i}}{\kappa_j+1}}\epsilon_j\right)=&\sum_{m=0}^{\infty}\Phi_j^{[m]}\left({\rm e}^{\frac{2\pi {\rm i}}{\kappa_j+1}}\epsilon_j\right)^{m},  \\
      \vdots&  \\
     \Phi_j\left({\rm e}^{\frac{2\kappa_j\pi {\rm i}}{\kappa_j+1}}\epsilon_j\right)=&\sum_{m=0}^{\infty}\Phi_j^{[m]}\left({\rm e}^{\frac{2\kappa_j\pi {\rm i}}{\kappa_j+1}}\epsilon_j\right)^{m}.
  \end{split}
\end{equation*}
It is readily to see that ${\rm e}^{\frac{2l_j\pi {\rm i}}{\kappa_j+1}}$ is a root for equation
$\mu^{\kappa_j+1}=1$. Moreover, it satisfies the equation $1+\mu+\cdots+\mu^{\kappa_j}=0.$
Denote $\mu_j\equiv {\rm e}^{\frac{2\pi {\rm i}}{\kappa_j+1}}$, then
\begin{equation*}
\begin{split}
   \Phi_{j[1]}\equiv \frac{1}{(\kappa_j+1)\epsilon_j^{\kappa_j}}\sum_{n=0}^{\kappa_j}\mu_j^{n}\Phi_j(\mu_j^n\epsilon_j) & =\sum_{m=0}^{\infty}
  \Phi_j^{[m(\kappa_j+1)+\kappa_j]}\epsilon_j^{m(\kappa_j+1)}, \\
    \Phi_{j[2]}\equiv\frac{1}{(\kappa_j+1)\epsilon_j^{\kappa_j-1}}\sum_{n=0}^{\kappa_j}\mu_j^{2n}\Phi_j(\mu_j^n\epsilon_j) & =\sum_{m=0}^{\infty}
  \Phi_j^{[m(\kappa_j+1)+(\kappa_j-1)]}\epsilon_j^{m(\kappa_j+1)}, \\
  \vdots&\\
  \Phi_{j[\kappa_j]}\equiv\frac{1}{(\kappa_j+1)\epsilon_j}\sum_{n=0}^{\kappa_j}\mu_j^{\kappa_jn}\Phi_j(\mu_j^n\epsilon_j) & =\sum_{m=0}^{\infty}
  \Phi_j^{[m(\kappa_j+1)+1]}\epsilon_j^{m(\kappa_j+1)},\\
  \Phi_{j[\kappa_j+1]}\equiv\frac{1}{\kappa_j+1}\sum_{n=0}^{\kappa_j}\Phi_j(\mu_j^n\epsilon_j) & =\sum_{m=0}^{\infty}
  \Phi_j^{[m(\kappa_j+1)]}\epsilon_j^{m(\kappa_j+1)}.
\end{split}
\end{equation*}
Then we choose the general solution:
\begin{equation*}
\begin{split}
     \Psi_j&=\sum_{l=1}^{\kappa_j+1}\alpha_l(\epsilon_j)\Phi_{j[l]} \\
    &=\frac{1}{(\kappa_j+1)\epsilon_j^{\kappa_j+1}}\sum_{n=0}^{\kappa_j}\left[\left(\sum_{l=1}^{\kappa_j+1}(\mu_j^{ n}\epsilon_j)^{l}\alpha_l(\epsilon_j)\right)
    \Phi_j(\mu_j^{n}\epsilon_j)\right]
\end{split}
\end{equation*}
where
\begin{equation*}
  \alpha_l(\epsilon_j)=\sum_{i=0}^{\infty}\alpha_l^{[i]}\epsilon_j^{(\kappa_j+1)i}.
\end{equation*}
In what following, we discuss how to choose the parameters. Assuming that the parameter $\alpha_{s_j}^{[0]}\neq 0$ is the first nonzero parameter in the list $\alpha_l^{[0]}$, $l=1,2,\cdots, \kappa_j+1,$ through the linear property for vector solutions we rewrite the vector function
\begin{equation*}
\begin{split}
 \Psi_j&=\sum_{l=1,l\neq s}^{\kappa_j+1}\widehat{\alpha_l}(\epsilon_j)\Phi_{j[l]},\,\, \widehat{\alpha_l}
  =\alpha_l/\alpha_s,\\
  &=\frac{\epsilon_j^{s_j}}{(\kappa_j+1)\epsilon_j^{\kappa_j+1}}\sum_{n=0}^{\kappa_j}\left[\left(1+\sum_{
  \substack{k=1,\\ k \mod\kappa_j+1\not\equiv 0}}^{\infty}\gamma_j^{[k]}(\mu_j^{ n}\epsilon_j)^{k}\right)
    \Phi_j(\mu_j^{n}\epsilon_j)\right].
\end{split}
\end{equation*}
A more convenient way to design the parameter is using the following expression
\begin{equation*}
  \exp\left[\sum_{\substack{k=1,\\ k \mod\kappa_j+1\not\equiv 0}}^{\infty}\beta_i^{[k]}(\mu_j^{n}\epsilon_j)^{k}\right]
\end{equation*}
to replace the expression
\begin{equation*}
  1+\sum_{
  \substack{k=1,\\ k \mod\kappa_j+1\not\equiv 0}}^{\infty}\gamma_j^{[k]}(\mu_j^{ n}\epsilon_j)^{k}.
\end{equation*}

Therefore, through the \textbf{proposition 1} we can obtain that the expansions
\begin{equation*}
  \frac{\Psi_k^{\dag}\Psi_j}{\widehat{\lambda_k}^*-\widehat{\lambda_j}}=\sum_{m=0,n=0}^{\infty,\infty}G_{k,j}^{[m+1,n+1]}\epsilon_k^{*m(\kappa_j+1)}\epsilon_j^{n(\kappa_k+1)},\,\, G_{k,j}^{[m,n]}=H_{k,j}^{[m(\kappa_j+1)-s_j,n(\kappa_k+1)-s_k]},
\end{equation*}
and
\begin{equation*}
  \frac{\Psi_k^{\dag}\Psi_j}{\widehat{\lambda_k}^*-\widehat{\lambda_j}}+\phi_k^*\psi_j^{[i]}=\sum_{m=0,n=0}^{\infty,\infty}G_{k,j}^{[i;m+1,n+1]}\epsilon_k^{*m(\kappa_j+1)}\epsilon_j^{n(\kappa_k+1)},\,\, G_{k,j}^{[i;m,n]}=H_{k,j}^{[i;m(\kappa_j+1)-s_j,n(\kappa_k+1)-s_k]},
\end{equation*}
where $\Psi_j=D_{N+1}\left[\phi_j,\psi_j^{[1]},\psi_j^{[2]},\cdots,\psi_j^{[N]}\right]^T.$ The parameters $s_i$ affect the structures and numbers of RWs. Some of results are exhibited in the previous studies \cite{lingzhao,ling-zhao-guo}. For the two component NLSE, if we choose the determining equation \eqref{cubic} with triple root, the first order RWs with the parameter $s_1=1$ correspond to two fundamental RWs and the second ones yield the six fundamental RWs; but for the parameter $s_1=2$, the first order RWs merely yield one fundemental RW and the second order RWs only involve four fundamental RWs \cite{lingzhao}.

To obtain the general multi-high-order RW solution, we merely need to take limit $\epsilon_j\rightarrow 0$ for the formula \eqref{gene-soliton}. Combining with the above results, we arrive at:
\begin{thm}
The general RW solutions for VNLSE can be represented as
\begin{equation}\label{NRW}
      q_i[K]=a_i\left[\frac{\det(G^{[i]})}{\det(G^{[0]})}\right]\exp{\theta_i},
\end{equation}
where
\begin{equation*}
  G^{[m]}=\begin{bmatrix}
     G_{1,1}^{[m]} & G_{1,2}^{[m]}& \cdots & G_{1,p}^{[m]} \\
     G_{2,1}^{[m]} & G_{2,2}^{[m]}& \cdots & G_{2,p}^{[m]} \\
     \vdots & \vdots & &\vdots \\
     G_{p,1}^{[m]} & G_{p,2}^{[m]} & \cdots & G_{p,p}^{[m]} \\
   \end{bmatrix},\,\, \left\{
                        \begin{array}{ll}
                          G_{k,j}^{[0]}=\left(G_{k,j}^{[r,s]}\right)_{\substack{1\leq r\leq K_k\\
    1\leq s\leq K_j}}, &  \\
                          G_{k,j}^{[m]}=\left(G_{k,j}^{[m;r,s]}\right)_{\substack{1\leq r\leq K_k\\
    1\leq s\leq K_j}}, & m\geq 1,
                        \end{array}
                      \right.
\end{equation*}
and $K=K_1+K_2+\cdots+K_p.$
\end{thm}
The generalized form can be used to derive RW solution with arbitrary order without the constrain conditions on background fields.  The solution formulas are given by a purely algebraic way. Especially, high-order RW with different fundamental patterns can be obtained, in contrast to the ones reported before \cite{Guo,lingzhao}. Since the spectral parameter admit more freedom than the case in scalar NLSE system, we can obtain mainly three cases for high-order RW solutions. The first case is iterating RW solution with different spectral parameters, which corresponds to superposition of fundamental RWs with different structures. The second case is iterating RW solution with an identical spectral parameter, which describes superpositions  of fundamental RWs with identical pattern. The third case is iterating RW solution by combining the previous two cases (iterate some steps with fixed spectral parameter and iterate some steps with different other spectral parameters), which admits superpositions of RWs with different patterns and identical pattern. RW solutions in the first case is called by multi-RW solutions, and the ones in the second case is called by high-order RW solutions, the ones in the third case is called by multi-high-order RW solutions.

\subsection{Classification of Fundamental rogue wave}

The fundamental RW solution can be given directly by the above formulas \eqref{NRW}. The solution can be presented as follows by some simplifications
\begin{equation}\label{fundrw}
q_i[1]=
a_i\left[1-\frac{1}{(p_1+b_i)^2+r_1^2}\frac{-2{\rm i}(p_1+b_i)(x+p_1t)+2{\rm i}r_1^2t+1}{(x+p_1t)^2+r_1^2t^2+\frac{1}{4r_1^2}}\right]{\rm e}^{\theta_i},\\
\end{equation}
where $p_1=\mathrm{Re}(\chi)$, $r_1=\mathrm{Im}(\chi)$ and $\chi$ is a double root for cubic equation \eqref{cubic}.
We find that there are three different types of rogue wave solution:
\begin{itemize}
  \item If $\frac{(p_1+b_i)^2}{r_1^2}\geq 3$, then the rogue wave is called anti-eye-shaped rogue wave (or dark RW).
  \item If $\frac{1}{3}< \frac{(p_1+ b_i)^2}{r_1^2}<3$, then the rogue wave is four-petaled rogue wave.
  \item If $\frac{(p_1+ b_i)^2}{r_1^2}\leq\frac{1}{3}$, then the rogue wave is called eye-shaped rogue wave (or bright RW).
\end{itemize}
Similar patterns for RW have been demonstrated in \cite{Chow}. The explicit conditions for the RW transition were not given there. Additionally, breathers with these different pattern units were derived in a three-component coupled system \cite{Zhao3}. The value of $\frac{(p_1+b_i)^2}{r_1^2}$ can be used to make judgment on the RW pattern in the two components conveniently  based on the above results. When the multi RWs interact with each other, there will be many different patterns for which it is hard to know which fundamental RWs constitute them. These criterions can be used to clarify the fundamental RW pattern conveniently. For example, we discuss a three-component case with
 \begin{equation}\label{para1}
  a_1=a_2=a_3=1,\,\, b_1=-b_3=1,\,\, b_2=0,
\end{equation}
then the equation \eqref{discrim} can be solved
\begin{equation*}
  \chi_1={\rm i},\,\,\chi_2=\frac{\sqrt{2}}{2}(1+{\rm i}),\,\,
  \chi_3= \frac{\sqrt{2}}{2}(-1+{\rm i}).
\end{equation*}
Inserting the above parameters into formula \eqref{fundrw}, we can obtain three types of different fundamental RW solutions. The reasons for existence of three different types of RWs can be demonstrated by the MI analysis in the following section \ref{sec5}.

\subsection{ Multi-rogue wave for VNLS}

Through above general RW solution \eqref{NRW}, we can obtain the multi-RW for the
$N$-component NLS equation by the different $b_i$. Actually, the cases of formula \eqref{NRW} are very complex: such as the degree of root maybe not two. For simplicity, we merely consider the degree of root is two.  From above general step, we can obtain that
\begin{equation*}
  E_{k,j}^{[0,0]}=1,\,\, E_{k,j}^{[0,1]}=\frac{1}{\chi_k^{[0]*}-\chi_j^{[0]}}
  ,\,\, E_{k,j}^{[1,0]}=-\frac{1}{\chi_k^{[0]*}-\chi_j^{[0]}},\,\,
  E_{k,j}^{[1,0]}=\frac{-2}{(\chi_k^{[0]*}-\chi_j^{[0]})^2}
\end{equation*}
and
\begin{equation*}
\begin{split}
    S_j^{[0]}&=1,\,\, S_j^{[1]}=(x+\chi_j^{[0]}t+\gamma_j),  \\
    S_k^{[0]}&=1,\,\, S_k^{[1]}=(x+\chi_k^{[0]}t+\gamma_k).
\end{split}
\end{equation*}
Ignoring the superscript $^{[0]}$,
and together with the above formula \eqref{NRW}, we can obtain the following multi-RW solution:
\begin{equation}\label{NRW1}
  q_i[p]=a_i\prod_{k=1}^{p}\frac{\chi_k^*+b_i}{\chi_k+b_i}\left[\frac{\det(F^{[i]})}{\det(F)}\right]{\rm e}^{\theta_i}
\end{equation}
where
\begin{equation*}
  \begin{split}
     F&=\left(F_{k,j}\right)_{1\leq k,j\leq p},\,\, F^{[i]}=\left(F_{k,j}^{[i]}\right)_{1\leq k,j\leq p}   \\
      F_{k,j}&=\frac{1}{\chi_k^{*}-\chi_j}\left[(x+\chi_jt+\gamma_j)
      (x+\chi_k^{*}t+\gamma_k^*)-\frac{{\rm i}(2x+(\chi_j+\chi_k^{*})t+\gamma_j+\gamma_k^*)}{\chi_k^{*}-\chi_j}
      -\frac{2}{(\chi_k^{*}-\chi_j)^2}\right],\\
       F_{k,j}^{[i]}&=\frac{1}{\chi_k^{*}-\chi_j}\left\{\left[{\rm i}(x+\chi_jt+\gamma_j)-\frac{1}{\chi_j+b_i}\right]
      \left[-{\rm i}(x+\chi_k^{*}t+\gamma_k^*)+\frac{1}{\chi_k^*+b_i}\right]\right.\\
      &\left.+\frac{-{\rm i}(2x+(\chi_j+\chi_k^{*})t+\gamma_j+\gamma_k^*)+\frac{1}{\chi_k^*+b_i}+\frac{1}{\chi_j+b_i}}{\chi_k^{*}-\chi_j}
      -\frac{2}{(\chi_k^{*}-\chi_j)^2}\right\}.
  \end{split}
\end{equation*}
$\gamma_i$s are complex parameters. The multi-RWs were firstly given by the algebraic geometry solution reduction method \cite{Kalla}. Here we reconstruct it by the Darboux dressing method.

\subsection{ High-order and multi-high-order rogue wave for VNLSE}

In this paragraph, we discuss the high-order RW and multi-high-order ones
for the vector NLSE. High-order rogue wave is a nonlinear superposition of rogue wave with an identical spectra parameter. Multi high-order rogue wave is a nonlinear superposition of multi rogue wave with the fixed different spectra parameters.  Actually, the abstract formula has been derived before. To illustrate
these formulas explicitly, here we give a specific example. We take the parameters as \eqref{para1}. The detailed algorithm can be stated as following:

\textbf{i) The parameters expansion}

The parameters expansions can be readily obtained by iteration algorithm. For an explicit example, choosing $a_1=a_2=a_3=1$ and $b_1=-b_3=1$, $b_2=0$, through the discriminant \eqref{discrim} we can obtain the branch points whose image part are greater than 0:
\begin{equation*}
  \chi_1={\rm i},\,\, \chi_2=\frac{\sqrt{2}}{2}(1+{\rm i}),\,\, \chi_3=\frac{\sqrt{2}}{2}(-1+{\rm i}).
\end{equation*}
The parameter $\chi_1={\rm i}$ corresponds the spectral parameter
\begin{equation*}
  \lambda_1=\chi_1-\frac{1}{\chi_1-1}-\frac{1}{\chi_1}-\frac{1}{\chi_1+1}=3{\rm i},
\end{equation*}
and both $\chi_2=\frac{\sqrt{2}}{2}(1+{\rm i})$ and $\chi_3=\frac{\sqrt{2}}{2}(-1+{\rm i})$ correspond the spectral parameter $\lambda_2=2\sqrt{2}{\rm i}$. It follows that
\begin{equation*}
\begin{split}
    \widehat{\lambda_1}=&3{\rm i}-\frac{{\rm i}}{2}\epsilon_1^2,  \\
    \widehat{\chi_1}=&{\rm i}+\epsilon_1-\frac{1}{2}{\rm i}{\epsilon_1}^{2}+\frac{1}{8}{\epsilon_1}^{3}-\frac{1}{4}{\rm i}{\epsilon_1}^{
4}+o({\epsilon_1}^{4})
\end{split}
\end{equation*}
and
\begin{equation*}
\begin{split}
    \widehat{\lambda_2}=&2\sqrt{2}{\rm i}-\sqrt{2}\epsilon_2^2,  \\
    \widehat{\chi_2}=&\left(1+{\rm i}\right)\frac{\sqrt {2}}{2}+\epsilon_2+ \left( -\frac{\sqrt {2}}{4}-
{\rm i}\frac{3\sqrt {2}}{4} \right) {\epsilon_2}^{2}- \left( 1-{\frac {9}{8}} {\rm i}
 \right) {\epsilon_2}^{3}+\frac{7}{4}\sqrt {2} {\epsilon_2}^{4}
+o({\epsilon_2}^{4})
\end{split}
\end{equation*}
and
\begin{equation*}
\begin{split}
    \widehat{\lambda_3}=&2\sqrt{2}{\rm i}+\sqrt{2}\epsilon_3^2,  \\
    \widehat{\chi_3}=&\left(-1+{\rm i}\right)\frac{\sqrt {2}}{2}+\epsilon_3+ \left(\frac{\sqrt {2}}{4}-
\frac{3\sqrt {2}}{4}{\rm i}\right) {\epsilon_3}^{2}- \left( 1+{\frac {9}{8}} {\rm i}
 \right) {\epsilon_3}^{3}-\frac{7}{4}\sqrt {2} {\epsilon_3}^{4}+o({\epsilon_4}^{4})
\end{split}
\end{equation*}

\textbf{ii) The coefficient of element}

Then we have the following expansions:
\begin{equation*}
\begin{split}
    \frac{\exp(\widehat{X_k}^*+\widehat{X_j})}{\widehat{\chi_k}^*-\widehat{\chi_j}}=&\sum_{m=0,n=0}^{\infty,\infty}
 G_{k,j}^{[0;m,n]}\epsilon_k^{*m}\epsilon_j^{n} \\
    \frac{\widehat{\chi_k}^*+b_i}{\widehat{\chi_j}+b_i}\frac{\exp(\widehat{X_k}^*+\widehat{X_j})}{\widehat{\chi_k}^*-\widehat{\chi_j}}=& \sum_{m=0,n=0}^{\infty,\infty}
 G_{k,j}^{[i;m,n]}\epsilon_k^{*m}\epsilon_j^{n}
\end{split}
\end{equation*}
where
\begin{equation*}
   \widehat{X_k}={\rm i}\left[(\widehat{\chi_k}-\chi_k)(x-x_k+p_k(\epsilon_k))+\frac{1}{2}(\widehat{\chi_k}^2-\chi_k^2)(t-t_k)\right]
\end{equation*}
and $x_k$,$t_k\in \mathbb{R}$, $p_k(\epsilon_k)=\sum_{s=0}^{\infty}p_k^{[s]}\epsilon_k^{2s},$ $p_k^{[s]}\in \mathbb{C}$, $k,j=1,2,3.$

\begin{figure}[tbh]
\centering
\includegraphics[height=50mm,width=150mm]{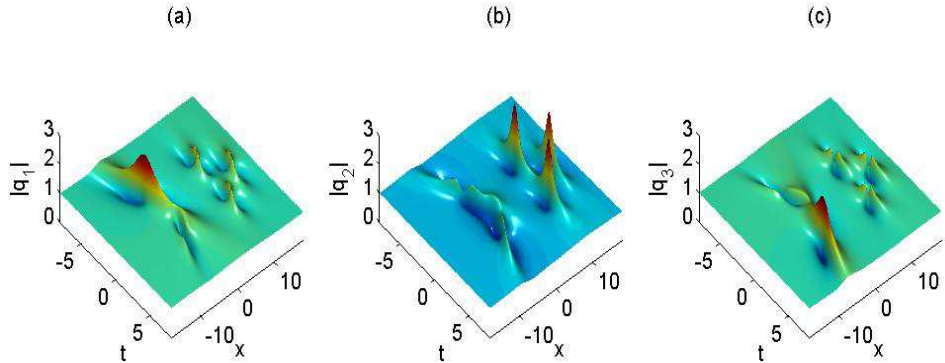} \hfil
\caption{(color online): The 3-D plot for multi-high-order rogue wave in a three-component case. (a) for $|q_1|$, (b) $|q_2|$ and (c) for $|q_3|$.}
\label{fig3nlsrw}
\end{figure}

For instance, the nonlinear superposition with second-order RW and two different
fundamental RW can be constructed as the following:
\begin{equation*}
  q_i[4]=\frac{\det(G^{[i]})}{\det(G^{[0]})}\exp(\theta_i)
\end{equation*}
where
\begin{equation*}
     G^{[m]}=\begin{bmatrix}
          G_{1,1}^{[m;1,1]} & G_{1,1}^{[m;1,3]}& G_{1,2}^{[m;1,1]} & G_{1,3}^{[m;1,1]} \\
          G_{1,1}^{[m;3,1]} & G_{1,1}^{[m;3,3]} & G_{1,2}^{[m;3,1]}& G_{1,3}^{[m;3,1]} \\
          G_{2,1}^{[m;1,1]} & G_{2,1}^{[m;1,3]} & G_{2,2}^{[m;1,1]} & G_{2,3}^{[m;1,1]} \\
          G_{3,1}^{[m;1,1]} & G_{3,1}^{[m;1,3]} & G_{3,2}^{[m;1,1]} & G_{3,3}^{[m;1,1]} \\
        \end{bmatrix}.
\end{equation*}
As an example, choosing the parameters $x_1=\frac{21}{2}$, $x_2=x_3=\frac{\sqrt{2}}{2}$, $t_1=t_2=t_3=0$,
and $p_1(\epsilon_1)=-10\epsilon_1^2$, $p_2(\epsilon_2)=p_3(\epsilon_3)=0,$ then we can plot the figure (Fig. \ref{fig3nlsrw}). It is seen that a second order RW which exhibits the triangle temporal-spatial distribution pattern is coexisted with two RWs. For the more complex cases, we could construct the exact RW solutions through the above similar algorithm. Here we merely give the algorithm to construct the RW solutions and omit the discussion of classification of high order RW solutions.

\section{The quantitative relation between fundamental AB or RW and modulational instability analysis}\label{sec5}
An important correspondence
between unstable linear eigenmodes and the Floquet spectral data of the integrable Lax
pair for the CNLS system was established \cite{Macl}, similar to results of Forest and Lee \cite{Forest} for the
scalar NLS equation. Several developments of the scalar NLS theory remain open for
the CNLS system. We would like to establish a quantitative relation between MI analysis and fundamental structures of AB and RWs.
The linearized stability of the plane wave solution can be readily obtained by Fourier analysis. We can perturb the seed solution with the following way:
\begin{equation*}
  q_i=q_{i}[0](1+p_i(x,t)).
\end{equation*}
Keeping the linear term of $p_i(x,t)$, the linearized disturbance equations become
\begin{equation*}
  {\rm i}(p_{i,t}+b_ip_{i,x})+\frac{1}{2}p_{i,xx}+\sum_{l=1}^{N}(p_{l}+p_{l}^*)=0,
  \,\, i=1,2,\cdots,N.
\end{equation*}
Suppose the perturbations $p_i(x,t)$s are periodical in $x$ with the period $2L$, $-L< x\leq L.$ Thus $p_i$ has the Fourier expansion
\begin{equation*}
  p_i(x,t)=\frac{1}{2L}\sum_{k=-\infty}^{+\infty}\widehat{p_{i,k}}{\rm e}^{{\rm i}\mu_k x}
\end{equation*}
where $\mu_k=2\pi k/2L$, $\widehat{p_{i,k}}=\int_{-L}^{L}p_i(x,t){\rm e}^{{\rm i}\mu_k x}\mathrm{d}x.$ Since the PDE is linear, it is sufficient to consider
\begin{equation}\label{perburb}
  p_i(x,t)=\widehat{p_{i,-k}}{\rm e}^{-{\rm i}\mu_k x}+\widehat{p_{i,k}}{\rm e}^{{\rm i}\mu_k x}
\end{equation}
for $k\neq 0$, while for $k=0$, then
\begin{equation*}
  p_i(x,t)=\widehat{p_{i,0}(t)}.
\end{equation*}
With the above analysis, we can obtain the modulational instability analysis for the
periodical perturbation. We give the judging criteria for $k\neq 0$. Firstly, we set
\begin{equation}\label{perburb1}
  \widehat{p_{i,k}}={\rm e}^{{\rm i}\mu_k\Omega_k t}p_{i,k},\,\,
  \widehat{p_{i,-k}}={\rm e}^{{\rm i}\mu_k\Omega_k^* t}p_{i,-k}^*.
\end{equation}
Then we can obtain the following equation
\begin{equation*}
  KY=0
\end{equation*}
where
\begin{equation*}
  \begin{split}
     K=&\mathrm{diag}\left((-\Omega_k-b_1-\frac{1}{2}\mu_k)\mu_k,\,
     (-\Omega_k+b_1-\frac{1}{2}\mu_k)\mu_k,\, \cdots,\,(-\Omega_k-b_N-\frac{1}{2}\mu_k)\mu_k,\,
     (-\Omega_k+b_N-\frac{1}{2}\mu_k)\mu_k\right)\\
     &+L\left(|a_1|^2,|a_1|^2,\cdots,|a_N|^2,|a_N|^2\right)  \\
     L=&\left(1,\,1,\, \cdots,\, 1,\, 1\right)^{\mathrm{T}}, \,\,
     Y=\left(p_{1,k},\,p_{1,-k},\, \cdots,\, p_{N,k},\, p_{N,-k}\right)^{\mathrm{T}}.
  \end{split}
\end{equation*}
Moreover, it is readily to obtain the dispersion relation:
\begin{equation*}
  1+\sum_{l=1}^{N}\frac{a_l^2}{(\Omega_k+b_l)^2-\frac{1}{4}\mu_k^2}=0.
\end{equation*}
From above assumption, we set $\alpha=\mu_k$, then we can deduce that the parameter $\chi+\frac{\alpha}{2}$ satisfies the same equation with $\Omega_k.$ Thus we establish the corresponding between the MI and AB solution quantitatively \eqref{charaN3}. This indicates that the structure of fundamental AB unit is determined by the MI dispersion forms.

For $k=0$, then we can obtain that
\begin{equation*}
  \begin{split}
     \frac{\mathrm{d}}{\mathrm{d}t}\widehat{p_{i,0R}}=&0, \\
     \frac{\mathrm{d}}{\mathrm{d}t}\widehat{p_{i,0I}}=&2\sum_{l=1}^{N}a_l^2\widehat{p_{i,0R}},
  \end{split}
\end{equation*}
here the subscripts $_R$ and $_I$ represent the real and image part respectively.
It is obviously that $\widehat{p_{i,0}}=\alpha_i+{\rm i}\beta_i t$, $\alpha_i$ and $\beta_i$ are some undetermined real parameters. Thus we know that this perturbation is instability. A simple way to avoid this instability is choosing the perturbation as $\int_{-L}^{L}p(x,0)\mathrm{d}x=0.$

To study the localized perturbation, we use the limit technique through taking limit $\mu_k\rightarrow 0$, i.e. $L\rightarrow \infty$.
Then the above Fourier series becomes Fourier transformation. However, to establish the relation with the rogue wave solution, we still use the denotation of Fourier series. If we take the limit $\mu_k\rightarrow 0$, then the dispersion relation can be represented as
\begin{equation*}
  1+\sum_{l=1}^{N}\frac{a_l^2}{(\Omega_k+b_l)^2}=0
\end{equation*}
which is nothing but the condition of RW. This indicates that the structure of fundamental RW is also determined by the MI dispersion forms. Explicitly, the dispersion form for
perturbations $\Omega_k$ can be used to know $\chi_R$ and $\chi_I$ directly
for certain backgrounds, namely $\chi=\Omega_k$. This is an equation correspondence, which is distinctive from the inequation correspondence for RW existence condition and baseband MI \cite{Baronio1}. The value of $\frac{(\chi_R+ b_i)^2}{\chi_I^2}$ can be used to make judgment on RW pattern in NLSE described systems conveniently. Obviously, if $\mathrm{Im}[\Omega_k]=0$, there will be no MI character, and the corresponding $\chi_I=0$ which brings RW solution meaningless. This character agrees well with the MI mechanism for RW in previous studies \cite{Baronio1,zhaoling}.  In this way, linear stability
analysis on plane wave backgrounds provide us a direct information for RW pattern.

This enables us to explain why scalar NLSE always admit eye-shaped pattern for
fundamental RW or AB unit. For scalar NLSE, there is only one component for which the background amplitude denoted by $a$ and the wave vector denoted by $b$. Then the dispersion relation give us the dispersion form $\Omega_k=-b+a\ {\rm i}=\chi$. The $\frac{(\chi_R+ b)^2}{\chi_I^2}=0$ always be smaller than $\frac{1}{3}$, which means that the fundamental RW for scalar NLSE is always an eye-shaped one. Similar calculations can be used to explain why anti-eye-shaped RW and four-petaled RW exist for vector RWs and AB units \cite{Zhao3,Degasperis,Chen3}.
The previous studies indicated that RW just exist in the cases with  $\mathrm{Im}[\Omega_k] \neq 0$. But the RW pattern must be observed numerically or experimentally. Here we report
 that the $\mathrm{Re}[\Omega_k]$ and $\mathrm{Im}[\Omega_k]$ can provide the fundamental RW pattern or AB unit type directly on certain backgrounds for NLSE described systems. The results allow one to predict RW or AB pattern and even the number of different fundamental RWs from linear stability analysis on plane wave background.

\section{Test MI dispersion form determining AB and RW pattern numerically}\label{sec6}
\begin{figure}[tbh]
\centering
\includegraphics[height=60mm,width=150mm]{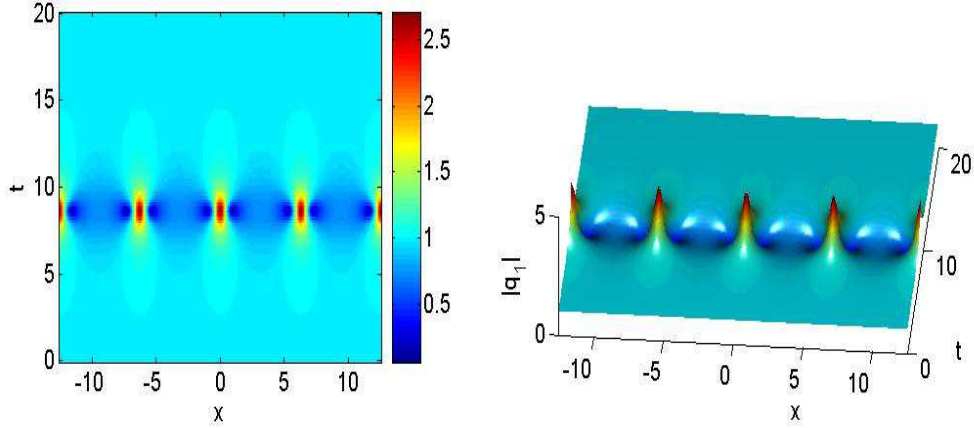} \hfil
\caption{(color online): The numerical simulation for $|q_1|$ with a special initial data. The left figure is the density plot, and the right one is the 3-D plot.}
\label{fign1}
\end{figure}

The above results suggest MI dispersion form determines the structure of fundamental AB or RW , which is supported by exact analytical solutions. However, AB or RW is a type of nonlinear excitation, which can be also excited from other initial conditions except the ones given by exact analytical solutions. As shown in \cite{Toenger}, many different initial periodic perturbations on plane wave background can evolve to be AB excitation. Therefore, we test the MI predictions about other initial conditions which are different from the ideal ones given by exact solutions. The discussions will make the theoretical results to be checked in real experiments. Through the asymptotical analysis, we obtain that
\begin{itemize}
  \item If $t\rightarrow -\infty$, then
\begin{equation}\label{aysm-}
  q_i=a_i {\rm e}^{\theta_i}\frac{\chi_{2}^*+b_i}{\chi_{2}+b_i}
\left[1+B_{i,-}{\rm e}^{\alpha \mathrm{Im}(\chi_{1})t}+o({\rm e}^{\alpha \mathrm{Im}(\chi_{1})t})\right]
\end{equation}
  \item If $t\rightarrow +\infty$, then
\begin{equation*}
  q_i=a_i {\rm e}^{\theta_i}\frac{\chi_{1}^*+b_i}{\chi_{1}+b_i}
\left[1+B_{i,+}{\rm e}^{-\alpha \mathrm{Im}(\chi_{1})t}+o({\rm e}^{-\alpha \mathrm{Im}(\chi_{1})t})\right]
\end{equation*}
\end{itemize}
where
\begin{equation*}
\begin{split}
    B_{i,-}&=\frac{\alpha(\chi_{2}^*-\chi_{2}){\rm e}^{-A_0}}{(\chi_{1}+b_i)(\chi_{2}^*-\chi_{1})}-\frac{\alpha(\chi_{2}^*-\chi_{2}){\rm e}^{A_0}}{(\chi_{2}^*+b_i)(\chi_{1}^*-\chi_{2})},  \\
    B_{i,+}&=\frac{\alpha(\chi_{1}^*-\chi_{1}){\rm e}^{-A_0}}{(\chi_{1}^*+b_i)(\chi_{2}^*-\chi_{1})}-\frac{\alpha(\chi_{1}^*-\chi_{1}){\rm e}^{A_0}}{(\chi_{2}+b_i)(\chi_{1}^*-\chi_{2})}, \\
    A_0&={\rm i}\alpha\left[x+\left(\mathrm{Re}(\chi_{1})+\frac{\alpha}{2}\right)t\right].
\end{split}
\end{equation*}
From above asymptotical expansion, we can replace the appropriate initial data with the precise initial data, to yield AB solution. For example, if we take $N=1$, through scaling and Galileo transformation, we can set $a_1=1$, and $b_1=0$. It follows that $\chi_1=-\frac{\alpha}{2}+{\rm i}\sqrt{1-\frac{\alpha^2}{4}}$, and $\chi_2=\frac{\alpha}{2}+{\rm i}\sqrt{1-\frac{\alpha^2}{4}}\equiv {\rm e}^{{\rm i}\gamma}$. Moreover, we can obtain that
\begin{equation*}
  B_{1,-}=\frac{\alpha(\chi_{2}^*-\chi_{2}){\rm e}^{-A_0}}{\chi_{1}(\chi_{2}^*-\chi_{1})}-\frac{\alpha(\chi_{2}^*-\chi_{2}){\rm e}^{A_0}}{\chi_{2}^*(\chi_{1}^*-\chi_{2})}=2\alpha\sqrt{1-\frac{\alpha^2}{4}}{\rm e}^{{\rm i}\gamma}\sin(\alpha x-\gamma).
\end{equation*}
Choosing $\alpha=1$, and the initial data $q_1(x,0)=(1+0.001{\rm e}^{\frac{\pi}{3}{\rm i}}\cos(x)){\rm e}^{-2{\rm i}}$, using the integrating-factor method \cite{yangbook}, we show that the dynamics is consistent with AB solution \cite{dega-book} (Fig. \ref{fign1}).

\begin{figure}[tbh]
\centering
\includegraphics[height=100mm,width=150mm]{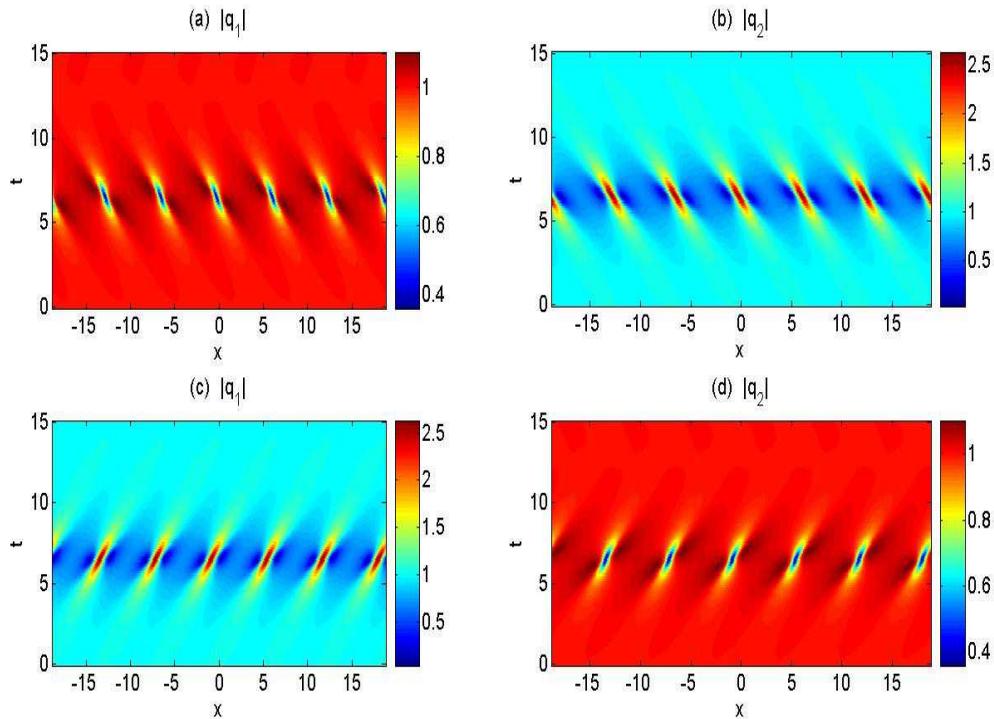} \hfil
\caption{(color online): The numerical test for two-component NLSE.}
\label{fign2}
\end{figure}

\begin{figure}[tbh]
\centering
\includegraphics[height=60mm,width=150mm]{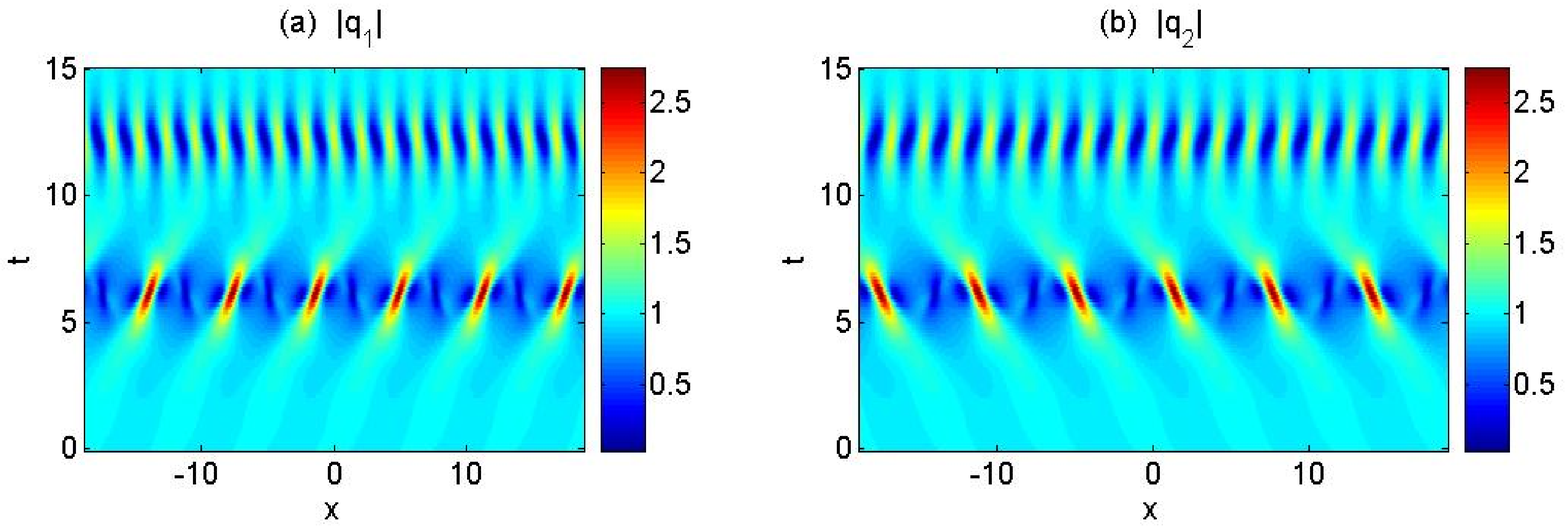} \hfil
\caption{(color online): The densityplot for $|q_1|$ and $|q_2|$.}
\label{fign3}
\end{figure}

When $N=2$, we take the initial data
\begin{equation*}
  \begin{split}
     q_1(x,0)=&\left[1+{\rm i}\epsilon\left(\frac{{\rm e}^{-{\rm i}x}}{(\chi_1+b_1)(\chi_2^*-\chi_1)}-\frac{{\rm e}^{{\rm i}x}}{(\chi_2^*+b_1)(\chi_1^*-\chi_2)}\right)\right]{\rm e}^{{\rm i}(x-3)}  \\
     q_2(x,0)=&\left[1+{\rm i}\epsilon\left(\frac{{\rm e}^{-{\rm i}x}}{(\chi_1+b_2)(\chi_2^*-\chi_1)}-\frac{{\rm e}^{{\rm i}x}}{(\chi_2^*+b_2)(\chi_1^*-\chi_2)}\right)\right]{\rm e}^{{\rm i}(-x-3)}
  \end{split}
\end{equation*}
where $\chi_1=0.4181886140+0.7701105965{\rm i}$, $\chi_2=\chi_1+1$, $b_1=-b_2=1$, $\epsilon=0.01$. Using the numerical method, we exhibit the dynamics which is also consistent with AB solution (Fig. \ref{fign2}(a,b)).
With the similar way, we can obtain another type AB solution (Fig. \ref{fign2}(c,d)) with the parameter $\chi_1=-1.4181886140+0.7701105965{\rm i}$, $\chi_2=\chi_1+1$.
If we take the periodical initial data
\begin{equation*}
  q_1(x,0)=(1+\epsilon\cos(x)){\rm e}^{{\rm i}(x-3)},\,\,
  q_2(x,0)=(1+\epsilon\cos(x)){\rm e}^{{\rm i}(-x-3)},
\end{equation*}
we obtain the dynamics for the figure (Fig. \ref{fign3}). It is seen that two distinct structures appear in the same time.

Finally,  we can see the perturbation \eqref{perburb} and \eqref{perburb1} in MI analysis process corresponds to the asymptotical form $B_{i,-}$ (\eqref{aysm-}) of AB solution precisely. Namely, the asymptotical analysis \eqref{aysm-} as $t\rightarrow -\infty$ illustrates that the MI analysis is consistent with AB solutions well. This completely explain why the quantitative correspondence between MI and fundamental AB can be obtained. Furthermore, the perturbation condition of RW can be approached well from AB's perturbation with period tend to be infinity. This explains completely why a quantitative correspondence is obtained between MI dispersion form and RW pattern.

\section{Conclusion and discussion}\label{sec7}
In this paper, we show that AB and RW pattern are determined by the dispersion form in MI regime. Moreover, we present a method to obtain multi- and high-order solutions to describe the nonlinear superpositions of these fundamental excitations. Based the solutions, many different types of complex excitations can be investigated exactly and analytically. This provides a possible way to predict the excitation pattern for them based on the dispersion form, without the need of solving the coupled nonlinear equations. These results will deepen and enrich our understanding and recognizing of AB and RW dynamics in many different nonlinear systems, such as nonlinear fiber, Bose-Einstein condensate, and plasma systems. The methods presented here can be applied to other nonlinear systems, such as three-wave resonant system \cite{thr-wav}, scalar NLSE with high-order effects \cite{hirota,he}, and other types coupled integrable models.

The MI analysis can be used to predict RW and AB pattern in $N$-component coupled system, which is supported by exact solutions in the above discussions. However, RW or AB can be also observed from many different initial perturbation conditions.
Recent studies indicate that RW comes from MI with resonance perturbations or baseband MI \cite{Baronio1,zhaoling}. Their excitation pattern does not depend on the profile
of initial perturbations, and it is determined by the MI characters of the nonlinear
 systems. Therefore, eye-shaped RW or breathers have been observed from many
 different initial conditions. Recently, dark RW (anti-eye-shaped RW) was
 observed in a real experiment and MI in vector system was demonstrated in a Manakov fiber system \cite{Frisquet2}. The results here have great
 possibilities to be checked in nonlinear fibers with two or more modes.

Kuznetsov-Ma (K-M) breather is another usual localized wave on plane wave background, which possesses resonant perturbation wave vector but different perturbation energy with the background \cite{zhaoling}. Its breathing period is determined by the perturbation energy difference. The mechanism of K-M breather will be studied in a separate literature paper, because the perturbation amplitude can be large which makes the linear stability analysis fail to explain its formation process.
For the general RW solutions which are derived in this paper, their amplifying process can be explained by the MI analysis. But as for the decaying process of these solutions, we can not find the mechanism up to now. It would be very interesting to explore the mechanism for the global dynamics of these solutions in the near future.

\section*{Acknowledgments}
This work is supported by National Natural Science Foundation of
China (Contact No. 11401221, 11405129)

\end{document}